\PassOptionsToPackage{table}{xcolor}

\documentclass[11pt]{article}

\usepackage[preprint]{acl2025}

\usepackage{times}
\usepackage{latexsym}

\usepackage{amsmath,amsfonts,bm}

\def\eqref#1{equation~\ref{#1}}

\def\1{\bm{1}}

\DeclareMathAlphabet{\mathsfit}{\encodingdefault}{\sfdefault}{m}{sl}
\SetMathAlphabet{\mathsfit}{bold}{\encodingdefault}{\sfdefault}{bx}{n}

\usepackage{hyperref}
\usepackage{url}
\usepackage{graphicx}
\usepackage{amsthm}
\theoremstyle{plain}

\theoremstyle{remark}

\usepackage{comment}
\usepackage{tcolorbox}
\usepackage[utf8]{inputenc}
\usepackage[T1]{fontenc}
\usepackage{booktabs}
\usepackage{amsfonts}
\usepackage{amssymb}
\usepackage{nicefrac}
\usepackage{microtype}
\usepackage{multirow}
\usepackage{tabularx}
\usepackage{wrapfig}
\usepackage{caption}
\usepackage{subcaption}
\usepackage{enumitem}
\usepackage{tikz}
\usetikzlibrary{calc, positioning, arrows.meta, shadows.blur, shadings, fit, backgrounds, decorations.pathreplacing, shapes.symbols, shapes.geometric}
\usepackage{pgfplots}
\pgfplotsset{compat=1.17}
\usepgfplotslibrary{colormaps}
\usepackage{amsmath}
\usepackage[capitalise,noabbrev]{cleveref}
\usepackage{xcolor}
\usepackage{array}
\usepackage{colortbl}
\definecolor{tabhdr}{RGB}{31,119,180}
\definecolor{tabrow}{RGB}{245,248,252}
\definecolor{yesgreen}{RGB}{34,139,84}
\definecolor{nored}{RGB}{190,60,60}

\definecolor{uscolor}{RGB}{31,119,180}
\definecolor{jpcolor}{RGB}{255,127,14}
\definecolor{twcolor}{RGB}{44,160,44}
\definecolor{krcolor}{RGB}{214,39,40}
\definecolor{hkcolor}{RGB}{148,103,189}

\definecolor{champion}{RGB}{244,246,250}        
\definecolor{rowwin}{RGB}{255,243,210}          
\definecolor{green(pigment)}{rgb}{0.0,0.65,0.31}
\definecolor{darksalmon}{rgb}{0.91,0.59,0.48}   

\usepackage{pifont}
\newcommand{\cmark}{\textcolor{green(pigment)}{\ding{51}}}
\newcommand{\xmark}{\textcolor{darksalmon}{\ding{55}}}
\tcbset{
  colback=blue!5,
  colframe=blue!40!black,
  coltitle=black,
  boxrule=0.8pt,
  arc=2mm,
  left=4pt, right=4pt, top=4pt, bottom=4pt
}

\renewcommand{\arraystretch}{1.05}
\usepackage{minitoc}        
\usepackage{graphicx}
\usepackage{xspace}
\usepackage{hyperref}   
\usepackage{fontawesome5} 

\IfFileExists{fontawesome5.sty}{%
  \usepackage{fontawesome5}
  
  \newcommand{\codeicon}{\faGithub}
  
}{%
  
  \newcommand{\codeicon}{Code}
  
}

\newcommand\blfootnote[1]{%
  \begingroup
  \renewcommand\thefootnote{}\footnote{#1}%
  \addtocounter{footnote}{-1}%
  \endgroup
}

\title{CrossAlpha: An Annual-Report Benchmark for Cross-Market \\Factor Research (with LLM Agents)}

\author{
  Qian Wang \quad
  Zhongyi Tong \quad
  Nuo Chen \quad
  Zhaomin Wu \quad
  Bingsheng He \\
  National University of Singapore \\
}

\begin{document}

\doparttoc
\faketableofcontents

\maketitle

\blfootnote{Correspondence to: Qian Wang $<$\href{mailto:qiansoc@nus.edu.sg}{qiansoc@nus.edu.sg}$>$.}

\begin{abstract}
Cross-market factor research studies whether firm-level signals from one or more markets can predict returns in a target market, but existing public benchmarks do not support cross-market disclosure-to-return evaluation. Building such a benchmark is challenging because filings differ across languages and regulatory systems, disclosure-derived similarity can be biased by common reporting components, and cross-market signals must be evaluated under feasible trading-time alignment. We introduce \textbf{CrossAlpha}, a public annual-report benchmark for cross-market factor research. CrossAlpha addresses these challenges through three corresponding components: \emph{Disclosure Distillation}, which standardises heterogeneous filings into ten-category English business descriptions; \emph{Residual Schema Graph Construction}, which builds PCA-whitened cross-market firm-pair scores from schema-level disclosures; and \emph{Timing-Aligned Evaluation}, which pairs the graph with 11 years of daily OHLCV data to construct forward-return labels under feasible cross-market execution protocols. CrossAlpha covers about 3,600 firms and 10,700 firm-year reports from the United States, Japan, Taiwan, South Korea, and Hong Kong, and releases about 19M directed firm-pair scores. In experiments, disclosure-derived cross-market peers outperform domestic text, industry-code, and return-correlation peers in the US-to-Japan setting (ICIR 0.39 versus 0.07--0.18), and cross-market sources beat the domestic text baseline in most target markets. CrossAlpha offers an open-sourced, reusable, return-grounded benchmark for cross-market financial NLP. \blfootnote{
\noindent
\raggedright
\urlstyle{same}
\codeicon\ \href{https://anonymous.4open.science/r/crossalpha-7346/}{anonymous.4open.science/r/crossalpha-7346/}
} 
\end{abstract}
\section{Introduction}
\label{sec:introduction}

\begin{figure}[t]
\centering
\includegraphics[width=\linewidth]{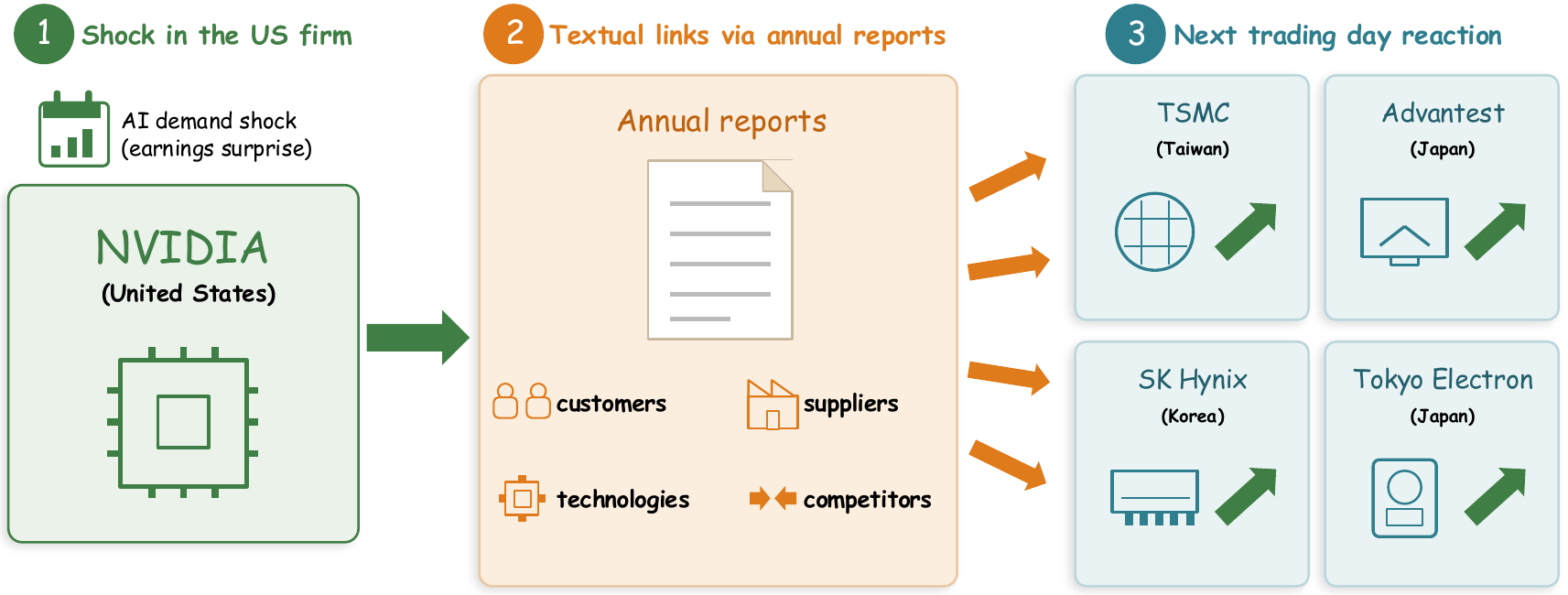}
\caption{Cross-market information flow. A return shock to a source firm transmits to foreign targets via relationships embedded in annual reports.}
\label{fig:intro_graph}
\end{figure}

In trading, a factor is a firm-level score used to rank stocks: a useful factor assigns higher scores to firms that subsequently earn higher market-relative returns~\citep{vuolteenaho2002drives}. Cross-market factor research asks whether signals from a \emph{source} market can rank stocks in a different \emph{target} market~\citep{menzly2010market, ma2019nonfinancial}. A target firm receives a high factor value when economically linked source-market firms have recently earned strong market-relative returns. Annual reports are ideal for this construction because they describe counterparties, technologies, and end-market exposures \emph{before} those relationships are fully priced into cross-market stock co-movement. As shown in \cref{fig:intro_graph}, an AI demand shock at NVIDIA can be translated into high factor scores for Asian semiconductor firms whose filings reveal related business exposures. Extracting cross-market alpha therefore presents a vital challenge for financial natural language processing (NLP): it requires mapping heterogeneous disclosures into a structural firm graph to capture these global return spillovers.

However, existing resources address only isolated parts of this problem. Financial NLP benchmarks provide standardized evaluation for language understanding and retrieval, but they rarely test whether text-derived firm links generate economically meaningful, forward-looking return predictions~\citep{shah2022flue,jorgensen2023multifin,finsts2024,tang2025finmteb}. Text-based peer methods derive useful firm similarity from filings, but they are mostly developed within a single market, ignoring global supply chains~\citep{hoberg2016text,chung2023modeling,setn2024,jagrivc2026ai}. Meanwhile, cross-market asset-pricing studies do evaluate return spillovers, but they typically rely on backward-looking price co-movement or statistically selected links rather than reusable, firm-level text~\citep{cspo2025,interintragnn2025,liu2026bipartite}. Thus, the field lacks a public benchmark that jointly connects multi-market disclosure text, dense economic links, aligned prices, and return-grounded factor evaluation.

To address these gaps, we introduce \textbf{CrossAlpha}, a public cross-market annual-report benchmark for cross-market factor research. Because capturing meaningful cross-border information flow requires economies with large, liquid traded universes and deep integration through global tech and manufacturing supply chains, CrossAlpha focuses on five heavily linked markets that meet these criteria: the United States, Japan, Taiwan, South Korea, and Hong Kong. 

Building a useful benchmark across these markets requires overcoming challenges at three distinct layers. First, at the \emph{disclosure layer}, raw annual reports suffer from severe filing heterogeneity across different languages and regulatory systems. Second, at the \emph{firm-link layer}, extracting genuine economic relationships is difficult because raw textual similarity is often biased by shared reporting-style components and broad industry jargon. Third, at the \emph{evaluation layer}, cross-market return prediction must respect strict temporal boundaries; mismatched time zones and trading calendars can easily introduce look-ahead bias if execution lags are not rigorously handled.

To address these challenges, we introduce a three-stage construction and evaluation pipeline. To resolve filing heterogeneity, an LLM-based \emph{Disclosure Distillation} stage converts the diverse annual reports into a shared ten-category English business schema. To mitigate similarity bias, a \emph{Residual Schema Graph Construction} stage applies PCA whitening to the schema embeddings, removing common reporting components to yield a dense, directed cross-market graph. Finally, to ensure rigorous backtesting, a \emph{Timing-Aligned Evaluation} protocol aligns the resulting semantic graph with 11 years of daily price data, supporting both monthly peer-momentum factors and dynamic LLM-agent event filtering under feasible execution lags.

Our contributions are summarized as follows:
\begin{itemize}[leftmargin=*]
\item We introduce CrossAlpha, an extensive public benchmark connecting standardized annual-report text, dense semantic firm-to-firm graphs, and 11 years of aligned daily prices across five major equity markets: the US, Japan, Taiwan, South Korea, and Hong Kong. It lays the groundwork for cross-market factor research by utilizing authentic economic relationships extracted from public corporate disclosures.
\item CrossAlpha consists of a reproducible Distiller--Graph--Linker pipeline and a market-aware evaluation protocol. This inclusive benchmark captures not only static structural firm links through monthly peer-momentum factors but also dynamic market reactions through high-frequency event-conditioned tasks, providing a more holistic view of global return spillovers.
\item Through detailed experiments, we demonstrate that disclosure-derived links markedly enhance cross-market return prediction. First, cross-market text peers significantly outperform domestic and non-text baselines, achieving up to a five-fold improvement in predictive skill (e.g., an ICIR of 0.39 in the US-to-Japan setting). Second, source selection reveals a directed \emph{Information Geography} that consistently yields positive long--short factors. Finally, using a GPT-5 agent to filter graph-retrieved neighbors dramatically improves event-driven spillover trading, nearly doubling the portfolio Sharpe ratio. 
\end{itemize}

\section{Related Work} \label{sec:related_work}
We discuss the most related work here and leave more details in \cref{app:related_detail}.

\begin{table*}[t]
\centering
\caption{Capability comparison for cross-market factor research. CrossAlpha is the only public resource that combines multi-market annual-report text, firm links, aligned prices, and a reusable return-prediction benchmark.}
\label{tab:dataset_comparison}
\small
\setlength{\tabcolsep}{3.5pt}
\renewcommand{\arraystretch}{1.10}
\resizebox{\linewidth}{!}{%
\begin{tabular}{@{}lccccccc@{}}
\toprule
\multirow{2}{*}{\textbf{Resource / Method}}
& \multicolumn{5}{c}{\textbf{Data}}
& \multicolumn{2}{c}{\textbf{Evaluation}} \\
\cmidrule(lr){2-6} \cmidrule(l){7-8}
& \textbf{\# Markets}
& \textbf{Firm text}
& \textbf{Firm links}
& \textbf{Prices}
& \textbf{Public}
& \textbf{Cross-market}
& \textbf{Return evaluation} \\
\midrule
\rowcolor{champion} TNIC \citep{hoberg2016text} & 1 & \cmark & \cmark & \xmark & \cmark & \xmark & \xmark \\
S-Peer \citep{chung2023modeling} & 1 & \cmark & \cmark & \cmark & \xmark & \xmark & \cmark \\
\rowcolor{champion} MD\&A Topic \citep{zhang2026uncovering} & 1 & \cmark & \cmark & \cmark & \xmark & \xmark & \cmark \\
ECC Topic \citep{jin2024business} & 1 & \cmark & \cmark & \cmark & \xmark & \xmark & \cmark \\
\rowcolor{champion} LLM-Augmented Semantic Network \citep{huang2026crossstock} & 1 & \cmark & \cmark & \cmark & \xmark & \xmark & \cmark \\
HybridFG \citep{hybridfirm2025} & 1 & \xmark & \cmark & \cmark & \xmark & \xmark & \cmark \\
\rowcolor{champion} US--CN BG \citep{liu2026bipartite} & 2 & \xmark & \cmark & \cmark & \xmark & \cmark & \cmark \\
CSPO \citep{cspo2025} & 2 & \xmark & \xmark & \cmark & \xmark & \cmark & \cmark \\
\rowcolor{champion} II-GNN \citep{interintragnn2025} & 5 & \xmark & \xmark & \cmark & \xmark & \cmark & \cmark \\
\midrule
\textbf{CrossAlpha (ours)} & \textbf{5} & \cmark & \cmark & \cmark & \cmark & \cmark & \cmark \\
\bottomrule
\end{tabular}}
\vspace{-6mm}
\begin{flushleft}
\footnotesize
\end{flushleft}
\end{table*}

\noindent \textbf{Comparison with Existing Resources.} \cref{tab:dataset_comparison} separates the related work by the capabilities needed for a reusable cross-market factor benchmark. TNIC~\citep{hoberg2016text} provides public 10-K-based firm links, but it is a single-market peer network rather than a return-prediction study. S-Peer~\citep{chung2023modeling}, MD\&A Topic~\citep{zhang2026uncovering}, ECC Topic~\citep{jin2024business}, and LLM-Augmented Semantic Network~\citep{huang2026crossstock} evaluate text-derived firm links through peer-return or cross-stock predictability, but they remain single-market studies and do not release a reusable cross-market benchmark. HybridFG~\citep{hybridfirm2025}, US--CN BG~\citep{liu2026bipartite}, CSPO~\citep{cspo2025}, and II-GNN~\citep{interintragnn2025} provide return-prediction evaluations using supply-chain, price, or learned trading-signal graphs, but not a public annual-report semantic graph with reusable disclosure text, aligned prices, and factor protocols. CrossAlpha is designed to fill exactly this missing intersection: public firm text, public firm links, aligned prices, explicit cross-market evaluation, and a reusable return-prediction protocol.

\section{CrossAlpha Construction}
\label{sec:construction}

\begin{figure*}[t]
\centering
\includegraphics[width=\textwidth]{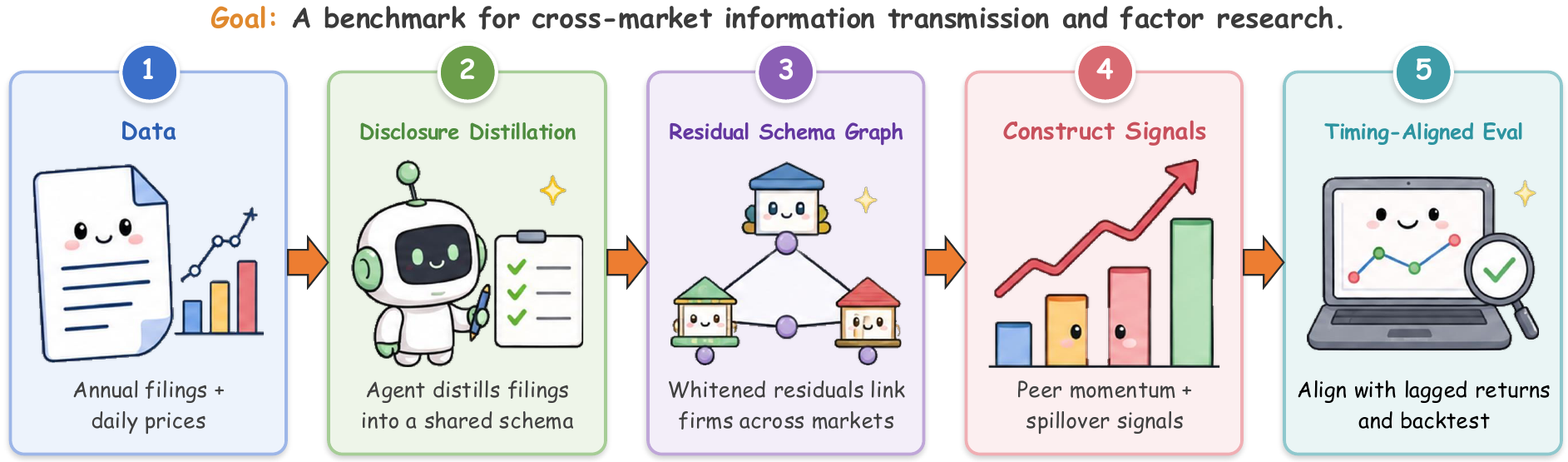}
\caption{CrossAlpha Overview. The framework standardises annual reports from multiple equity markets through Disclosure Distillation, constructs a Residual Schema Graph from the resulting schema-field representations, and aligns the graph with price data for timing-aware return-prediction benchmarks.}
\label{fig:overview}
\end{figure*}

We build CrossAlpha with two construction techniques, followed by timing-aligned return evaluation, as shown in \cref{fig:overview}. \emph{Disclosure Distillation} addresses filing heterogeneity by converting annual reports from five disclosure systems into a shared ten-category English business schema (\cref{sec:data_collection,sec:disclosure_distillation}; Appendix~\ref{app:schema_grounding}). \emph{Residual Schema Graph Construction} addresses common-component-biased similarity by embedding schema fields, applying PCA whitening to remove shared reporting-style components, and aggregating category-level residual similarities into a directed cross-market graph (\cref{sec:residual_schema_graph}; Appendix~\ref{app:graph_details}). The remaining timing challenge is handled by \emph{Timing-Aligned Evaluation} in the return-prediction protocols, where the graph is aligned with daily prices and tested under feasible execution lags.

\subsection{Data Collection}
\label{sec:data_collection}


\noindent \textbf{Textual Filings.} CrossAlpha covers five economically
significant and tightly connected equity markets: the United States
(Russell~1000, SEC EDGAR), Japan (TOPIX~500, EDINET), Taiwan (TWSE, MOPS),
South Korea (KOSPI, DART), and Hong Kong (HK Main Board). These markets provide
large listed-firm universes and dense US--Asia and intra-Asia trade,
technology, and supply-chain linkages, making them a useful testbed for
cross-market factor research. The official disclosure portals and their
usage-policy links are listed in \cref{app:data_sources}.

\noindent \textbf{Market Data.} To backtest the factors we extract from the cross-market dataset, we collect daily OHLCV (open, high, low, adjusted close, and volume) from Yahoo Finance\footnote{\url{https://finance.yahoo.com}} for January~2015--May~2026 across all five markets, together with historical shares outstanding for market-capitalisation neutralization.

\subsection{Disclosure Distillation}
\label{sec:disclosure_distillation}
\label{sec:feature_extraction}

We design the Distiller schema to convert heterogeneous annual reports into comparable business-exposure representations for cross-market firm-link construction. The schema satisfies two requirements: each category should capture a cross-firm channel that prior return-predictability work finds informative, and each category should be recoverable from mandatory annual filings across all five regulatory systems. The resulting ten-category English schema spans products and technologies, customers and supply chains, geographic and financial exposure, and business model, strategy, and competition. This design makes downstream similarity depend on comparable firm-level business content rather than market-specific filing formats. The category rationale and cross-regulator grounding are reported in \cref{tab:schema_rationale} and Appendix~\ref{app:schema_grounding}. 

Given LLMs' capability to read financial text and extract structured information reliably~\citep{lopezlira2023chatgpt,kim2024financial}, we instantiate the schema with an LLM-based Distiller: a single GPT-4.1~\citep{openai2025gpt41} pass per filing reads the document end-to-end and emits the standardised fields; the full prompt is given in \cref{app:distiller_prompt}.

\subsection{Residual Schema Graph Construction}
\label{sec:residual_schema_graph}
\label{sec:graph_construction}

\noindent \textbf{Residual Schema Encoding.}
We embed each standardised category text with OpenAI's
text-embedding-3-large~\citep{openai2024embedding} to obtain category-level
firm representations. However, even after standardisation, these embeddings
retain strong common components: all summaries are written in a similar
business-description style, which can pull unrelated firms closer in vector
space and inflate raw cosine similarities~\citep{su2021whitening}. Since
cross-market peer discovery should depend on firm-specific business exposure
rather than shared reporting style, we apply PCA whitening to the category
vectors before computing similarities~\citep{kessy2018optimal,su2021whitening}.
This transformation suppresses dominant common components and balances variance
across embedding directions, making the remaining similarities more reflective
of firm-specific schema content. We refer to the resulting PCA-whitened category
vectors as residual schema encodings.

\noindent \textbf{Directed Graph Construction.}
Using the residual schema encodings, CrossAlpha connects economically similar
firms across markets into a dense directed graph, which we call the Residual
Schema Graph:
\begin{equation}
\begin{gathered}
\mathcal{G}_{\mathrm{text}}=(V,E_{\mathrm{text}}),\qquad
V = \bigcup\nolimits_{m=1}^{K} V_{\mathcal{M}_m},\\
\mathcal{M}\in\{\mathrm{US},\mathrm{JP},\mathrm{TW},\mathrm{KR},\mathrm{HK}\},
\end{gathered}
\label{eq:text_graph}
\end{equation}
where nodes are listed firms partitioned by home market and edges represent
firm-to-firm residual schema similarities. For an ordered source market
$\mathcal{M}_b$ and target market $\mathcal{M}_a$, the similarity between a
target firm $i\in\mathcal{M}_a$ and a source firm $j\in\mathcal{M}_b$ is the
equal-weighted average of their ten category-level cosine similarities:
\begin{equation}
s_{\mathrm{text}}(i,j) = \frac{1}{10}\sum_{c=1}^{10}
\cos\!\left(\mathbf{z}_i^{(c)},\, \mathbf{z}_j^{(c)}\right),
\label{eq:text_sim}
\end{equation}
where $\mathbf{z}_i^{(c)}\in\mathbb{R}^{128}$ is the residual schema encoding of
firm $i$'s category-$c$ text. This construction yields a dense firm-to-firm peer
layer across all 25 directed market pairs.


\subsection{Dataset Statistics}
\label{sec:statistics}

\cref{tab:data_summary} highlights three properties of CrossAlpha. First, the benchmark provides broad market coverage: it spans 3{,}587 listed firms across the United States, Japan, Taiwan, South Korea, and Hong Kong, giving each target firm a large cross-market peer universe. Second, the released artifacts are aligned across modalities: the 10{,}700 firm-year schema records are paired with category-level representations, a dense firm-pair graph, and daily OHLCV data. Third, the graph layer expands the filing layer into about 19M directed firm-pair scores, providing the evaluation surface for cross-market peer selection and return prediction. Construction costs are in \cref{app:cost}.

\begin{table}[ht]
\centering
\caption{CrossAlpha dataset summary.}
\label{tab:data_summary}
\footnotesize
\setlength{\tabcolsep}{3.5pt}
\renewcommand{\arraystretch}{1.08}
\begin{tabularx}{\columnwidth}{@{}>{\raggedright\arraybackslash}p{0.28\columnwidth}
>{\raggedright\arraybackslash}X
>{\raggedleft\arraybackslash}p{0.20\columnwidth}@{}}
\toprule
\multicolumn{3}{@{}l}{\textit{Filing universe}} \\
\midrule
\textbf{Market} & \textbf{Universe} & \textbf{\# of Firms} \\
\midrule
\rowcolor{champion}
United States & Russell~1000 & 880 \\
Japan & TOPIX~500 & 453 \\
\rowcolor{champion}
Taiwan & TWSE & 1{,}040 \\
South Korea & KOSPI & 612 \\
\rowcolor{champion}
Hong Kong & Main board & 602 \\
\textbf{Total} & \textbf{5 equity markets} & \textbf{3{,}587} \\
\midrule
\multicolumn{3}{@{}l}{\textit{Benchmark components}} \\
\midrule
\textbf{Layer} & \textbf{Unit} & \textbf{Scale} \\
\midrule
\rowcolor{champion}
Schema text & ten-category firm-year records & 10{,}700 \\
Residual schema encodings & firm-category vectors & ${\approx}$107K \\
\rowcolor{champion}
Residual Schema Graph & directed firm-pair scores & ${\approx}$19M \\
Daily prices & firm-day OHLCV records & ${\approx}$10M \\
\rowcolor{champion}
\textbf{Total} & \textbf{schema, encoding, graph, and price records} & \textbf{${\approx}$29.1M} \\
\bottomrule
\end{tabularx}
\end{table}

\section{Timing-aligned Evaluation Protocol}
\label{sec:evaluation_protocol}
\label{sec:backtesting}

CrossAlpha serves not only as a textual and structural resource but also provides a standardized, return-grounded evaluation protocol. To isolate the predictive value of disclosure links, this protocol fixes the portfolio construction, varying only the source of firm links or the evaluation horizon.

\begin{figure}[t]
\centering
\resizebox{\columnwidth}{!}{%
\begin{tikzpicture}[font=\small,>=Stealth]
\node[align=center, font=\small] (hdr) {Target stocks\\ranked by factor};
\node[draw, fill=green!18, minimum width=1.8cm, minimum height=0.58cm, below=6pt of hdr] (q5) {Q5};
\node[draw, fill=gray!10, minimum width=1.8cm, minimum height=0.58cm, below=0pt of q5] (q4) {Q4};
\node[draw, fill=gray!10, minimum width=1.8cm, minimum height=0.58cm, below=0pt of q4] (q3) {Q3};
\node[draw, fill=gray!10, minimum width=1.8cm, minimum height=0.58cm, below=0pt of q3] (q2) {Q2};
\node[draw, fill=red!15, minimum width=1.8cm, minimum height=0.58cm, below=0pt of q2] (q1) {Q1};
\node[draw, rounded corners, fill=green!18, right=1.1cm of q5, align=center] (long) {Long\\Q5};
\node[draw, rounded corners, fill=red!15, right=1.1cm of q1, align=center] (short) {Short\\Q1};
\draw[->] (q5.east) -- (long.west);
\draw[->] (q1.east) -- (short.west);
\node[draw, rounded corners, fill=blue!10, right=4.3cm of q3, align=center] (ls) {Long--short\\book\\(Q5--Q1)};
\draw[->] (long.east) -- (ls.west);
\draw[->] (short.east) -- (ls.west);
\end{tikzpicture}}
\caption{Portfolio sort. Each month, target-market stocks are ranked by factor value and split into five equal-count quintiles. The long--short (LS) book buys the top quintile (Q5) and sells the bottom (Q1).}
\label{fig:portfolio_sort}
\end{figure}
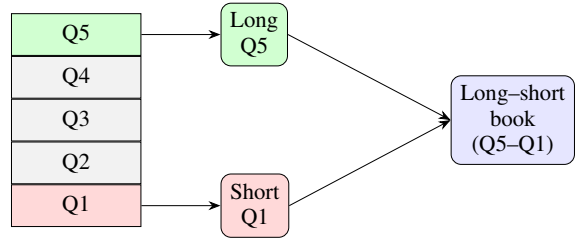

\noindent\textbf{Universe, portfolio, and timing.}
The released OHLCV panel spans January 2015--May 2026. The monthly factor tasks use January 2015--December 2025 with a strict one-month return lag: peer returns through month $M{-}1$ are used to form the month-$M$ target-stock portfolio. Crucially, the backtests use the released FY2024 text-similarity graph as a fixed benchmark input; return information enters solely through the lagged peer returns. For each source--target market pair, the traded universe is strictly the target market. Source-market firms only define peer sets and supply lagged returns, while target-market stocks are ranked into quintiles and held as an equal-weighted Q5--Q1 long--short book, as shown in \cref{fig:portfolio_sort}. While returns are strictly lagged to prevent price leakage, applying the static FY2024 graph to the historical panel evaluates the structural return-relevance of these textual links, rather than simulating an out-of-sample trading strategy (which would require point-in-time dynamic graphs).

\noindent\textbf{Factor Construction.}
For a target stock $i\in\mathcal{M}_a$ and source market $\mathcal{M}_b$, the
factor is the similarity-weighted average of source peers' sector-relative
returns,
\begin{equation}
f_i^{\mathrm{text}}(t) = \frac{\sum_{j \in \mathcal{M}_b} \alpha_{ij}^{\mathrm{text}} \cdot r_j^{(L)}(t)}{\sum_{j \in \mathcal{M}_b} \alpha_{ij}^{\mathrm{text}}},
\label{eq:factor}
\end{equation}
where $\alpha_{ij}^{\mathrm{text}}$ is the sigmoid top-1\% similarity weight
(\cref{eq:sigmoid}) and $r_j^{(L)}(t)$ is peer $j$'s $L$-month
sector-relative return. Baselines reuse this identical pipeline, altering only the peer definition; source-selection variants change only which source market supplies the peers.

\noindent\textbf{Metrics.}
We report six equity-factor metrics: (1) IC and (2) ICIR measure
rank-predictive skill~\citep{grinold2000active}, while the long--short
(3) Sharpe~\citep{sharpe1994}, (4) MaxDD, (5) Ret, and (6) CumRet summarize the
portfolio after a 2~bp one-way transaction cost. IC is the monthly Spearman rank correlation
between factor scores and next-month returns. ICIR is the annualized
mean-to-volatility ratio of that monthly IC series, following standard firm-link
return-predictability literature~\citep{cohen2008economic,menzly2010market}. Full
metric definitions and equations are provided in \cref{app:metrics,eq:ic,eq:icir}.

\noindent\textbf{Neutralization.}
For cross-market source comparisons—including source selection,
Cross-Market Gain, and Information-Geography—we report signed ICIR after
GICS-sector and size neutralization. We retain the sign to explicitly capture reverse predictability (negative values). This source-comparison metric aligns with recent
relational benchmarks~\citep{interintragnn2025,hybridfirm2025}, ensuring that we measure purely relational, within-sector and within-size skill rather than a latent sector or size tilt, with details in \cref{app:neutralization}.
\section{Experiments} \label{sec:experiments}
Using the timing-aligned protocol in \cref{sec:evaluation_protocol}, we evaluate
CrossAlpha through three benchmark questions:
\textbf{RQ1:} Do peers defined by cross-market filing text predict returns
better than peers from industry codes, price correlation, or a single market?
\textbf{RQ2:} Where should those peers come from, and is one source market best
for every target or does the best source differ by target? \textbf{RQ3:} Beyond
the monthly factor, does the same peer graph support event-conditioned daily
spillover prediction?

\subsection{RQ1: Cross-Market Text vs.\ Single-Market and Non-Text Peers}
\label{sec:rq1}


\noindent\textbf{Approach.}
We fix one source--target market pair: target Japan TOPIX~500, and source US
Russell~1000. All methods share the identical monthly peer-momentum pipeline and differ only in how peers
are defined. We compare CrossAlpha's cross-market text embedding against three alternatives:
(i)~a single-market text baseline (JP-to-JP domestic peers, identical
pipeline), which tests whether cross-market text is needed at all; (ii)~GICS
sector equal-weight peers; and (iii)~252-day rolling return correlation, the
standard non-text peer definition. Full baseline settings are in \cref{app:baselines}.

\begin{table}[t]
\centering
\caption{US-to-JP comparison. All rows share the same monthly pipeline and differ only in the peer definition.}
\label{tab:baselines_monthly}
\resizebox{\columnwidth}{!}{%
\begin{tabular}{@{}lrrrrrr@{}}
\toprule
\textbf{Peer definition} & \textbf{IC} & \textbf{ICIR} & \textbf{Sharpe} & \textbf{MaxDD} & \textbf{Ret} & \textbf{CumRet} \\
\midrule
\rowcolor{champion}
GICS sector & 0.006 & 0.18 & 0.25 & -14.3\% & 1.1\% & 11.6\% \\
Return correlation & 0.004 & 0.11 & 0.23 & -8.4\% & 1.0\% & 10.7\% \\
\rowcolor{champion}
JP-to-JP text & 0.003 & 0.07 & 0.05 & -9.5\% & 0.3\% & 3.0\% \\
\midrule
\textbf{US-to-JP text} & 0.009 & 0.39 & 0.39 & -6.8\% & 1.2\% & 13.2\% \\
\bottomrule
\end{tabular}%
}
\end{table}

\begin{figure}[t]
\centering
\includegraphics[width=0.92\linewidth]{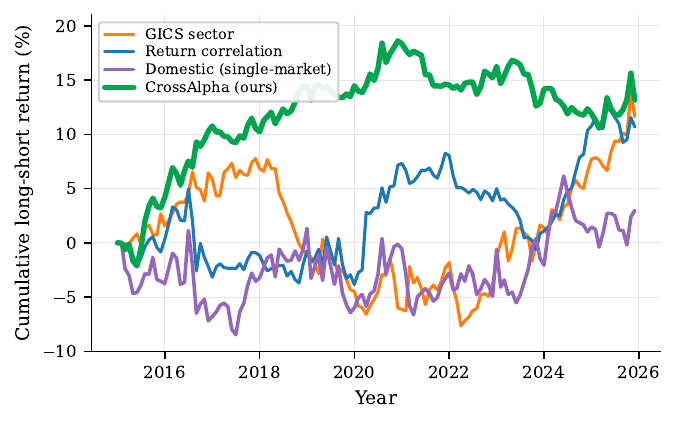}
\caption{Cumulative long-short return on US-to-JP under the identical monthly pipeline, with methods differing only in the peer definition.}
\label{fig:rq1_cumret}
\end{figure}

\cref{tab:baselines_monthly} reports the comparison and
\cref{fig:rq1_cumret} plots the cumulative long-short returns. We have two findings as follows:

\noindent \textbf{Cross-market text beats the single-market baseline.} Building
peers from domestic Japanese filings barely helps. JP-to-JP text peers reach an
ICIR of only 0.07, so same-market text similarity carries almost no predictive
signal. Drawing the same peers from the US instead lifts the ICIR to 0.39. We
attribute this gap to where the information originates rather than to the text
itself: investors price firms inside their own market almost immediately,
leaving no time gap for a domestic peer to exploit, whereas a US disclosure is
reflected in US prices first and reaches the linked Japanese firm only later,
and the factor profits from that
delay~\citep{menzly2010market}. \cref{fig:rq1_cumret}
makes the contrast visible: the cross-market portfolio compounds steadily and
stays above every baseline, while the domestic portfolio drifts around zero.

\noindent \textbf{Text also beats the non-text peer definitions, which break
cross-market.} GICS sector matching reaches an ICIR of 0.18 and 252-day return
correlation 0.11, against 0.39 for text, a two- to three-fold gap that also
shows up in risk (Sharpe 0.39, max drawdown -6.8\% versus
GICS's -14.3\% as shown in \cref{tab:baselines_monthly}). We attribute this to what
each definition measures: a GICS label is too coarse to say whether a US and a
Japanese firm are actually each other's customer, supplier, or competitor, while
return correlation is backward-looking and captures the result of information
flow rather than its cause. Business-description text is more direct in this
setting because it can identify firm-level economic links before the linked
returns are observed.

\subsection{RQ2: Source Selection and Directed Information Geography} \label{sec:rq2}

Because \cref{sec:rq1} shows in the US-to-JP setting that cross-market text beats
single-market and non-text peer definitions, we next ask whether that result is general for other markets. We therefore scale the same monthly factor construction to all five
markets and ask which source market gives the strongest factor for each target
and how directional those cross-market links are.

\noindent\textbf{Approach.} We extend the single US-to-JP source--target pair to every
ordered pair of markets. For each source $s$ and target $t$ in \{US, JP, TW, KR,
HK\} we build the same peer-momentum factor as in \cref{sec:rq1} and record signed
annualised ICIR as a compact predictive statistic for comparing
source-market choices. These scores form a 5-by-5 matrix $\mathbf{G}$ that we call the
\emph{Cross-Market Information-Geography Matrix}, where $\mathbf{G}_{st}$ is how
strongly source $s$ predicts target $t$. Reading a column gives each target's best single
cross-market source, and comparing $\mathbf{G}_{st}$ with $\mathbf{G}_{ts}$ gives
the direction of information flow.

Analyzing this Information-Geography Matrix reveals a complex network of global predictability. Specifically, we draw four key insights regarding how and where alpha signals propagate:

\begin{table}[t]
\centering
\caption{Cross-market generalization: US-to-T peer-momentum factors across four Asian targets.}
\label{tab:multi_market}
\scriptsize
\setlength{\tabcolsep}{2.8pt}
\begin{tabular}{@{}llrrrrrr@{}}
\toprule
\textbf{Market} & \textbf{LB} & \textbf{IC} & \textbf{ICIR} & \textbf{Sharpe} & \textbf{MaxDD} & \textbf{Ret} & \textbf{CumRet} \\
\midrule
\rowcolor{champion} JP & 6mo  & 0.008 & 0.31 & 0.30 & -8.2\% & 1.0\% & 10.8\% \\
JP & 12mo & 0.009 & 0.39 & 0.39 & -6.8\% & 1.2\% & 13.2\% \\
\midrule
\rowcolor{champion} TW & 6mo  & 0.008 & 0.41 & 0.71 & -5.9\% & 2.4\% & 26.6\% \\
TW & 12mo & 0.009 & 0.48 & 0.91 & -3.0\% & 2.9\% & 32.1\% \\
\midrule
\rowcolor{champion} KR & 6mo  & 0.009 & 0.46 & 0.49 & -8.3\% & 2.4\% & 25.6\% \\
KR & 12mo & 0.012 & 0.61 & 0.64 & -4.3\% & 5.6\% & 60.8\% \\
\midrule
\rowcolor{champion} HK & 6mo  & 0.015 & 0.66 & 0.98 & -7.1\% & 5.0\% & 54.1\% \\
HK & 12mo & 0.013 & 0.52 & 0.88 & -6.4\% & 4.7\% & 50.7\% \\
\bottomrule
\end{tabular}
\end{table}

\begin{figure}[t]
\centering
\includegraphics[width=0.9\columnwidth]{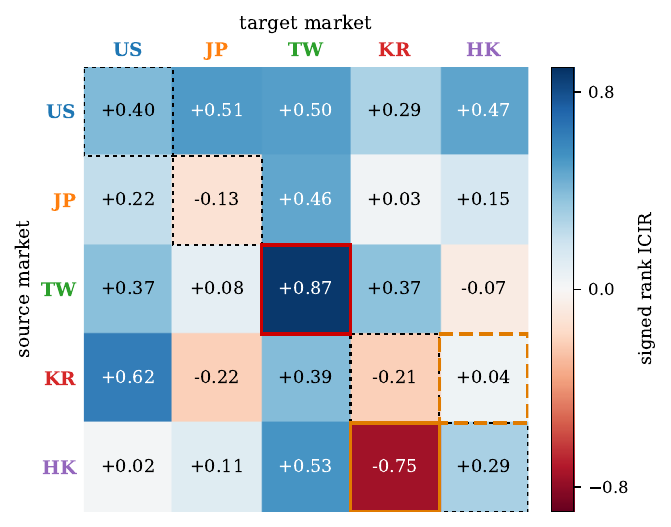}
\caption{Cross-Market Information-Geography Matrix $\mathbf{G}$. $(s,t)$ is the ICIR of the source-$s$-to-target-$t$ factor.}
\label{fig:infogeo_matrix}
\end{figure}

\begin{figure}[t]
  \centering
  \includegraphics[width=\columnwidth]{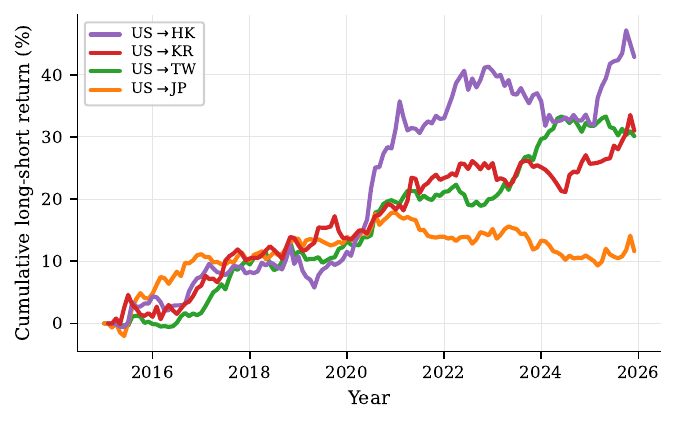}
  \caption{Cross-market generalization across four Asian targets.}
  \label{fig:rq2_generalization}
\end{figure}

\begin{table}[t]
\centering
\caption{Cross-market performance for each target market. The ICIR block
compares the cross-market signal from the source matrix with the same-market
baseline, with Gain defined as Cross-market minus Domestic.}
\label{tab:best_source}
\scriptsize
\setlength{\tabcolsep}{2.6pt}
\begin{tabular}{@{}lrrrrrr@{}}
\toprule
\textbf{Target} & \multicolumn{3}{c}{\textbf{ICIR}} & \multicolumn{3}{c}{\textbf{Portfolio}} \\
\cmidrule(lr){2-4}\cmidrule(l){5-7}
& \textbf{Cross-market} & \textbf{Domestic} & \textbf{Gain} & \textbf{Sharpe} & \textbf{MaxDD} & \textbf{Ret} \\
\midrule
\rowcolor{champion}
US & 0.62 & 0.40 & 0.22 & 0.19 & -5.8\% & 0.4\% \\
JP & 0.51 & -0.13 & 0.64 & 0.06 & -11.0\% & 0.3\% \\
\rowcolor{champion}
TW & 0.53 & 0.87 & -0.34 & 0.82 & -5.3\% & 2.5\% \\
KR & 0.37 & -0.21 & 0.58 & 0.06 & -26.2\% & 0.3\% \\
\rowcolor{champion}
HK & 0.47 & 0.29 & 0.18 & 0.82 & -5.4\% & 4.0\% \\
\bottomrule
\end{tabular}
\end{table}

\noindent\textbf{CrossAlpha finds positive cross-market factors across all five
targets.} As shown in \cref{tab:multi_market}, the benchmark turns
annual-report links into tradable cross-market signals. After selecting the best
non-domestic source for each target, the long--short factor is positive in every
market in \cref{tab:best_source}, with annualised returns from 0.3\% to 4.0\%
and Sharpe ratios up to 0.82. The fixed US-source experiment tells the same
story from a stricter angle: even when using only US peers as the source, the
factor performs strongly across Asian target markets.

\noindent\textbf{Source selection reveals target-specific information geography.}
The strongest non-domestic source changes with the target market: Korea is best for
the US, the US is best for Japan and Hong Kong, Hong Kong is best for Taiwan,
and Taiwan is best for Korea, as shown in
\cref{fig:infogeo_matrix}. Four different source markets appear
across five targets, so CrossAlpha exposes information geography that a
fixed-source benchmark would miss.

\noindent\textbf{Cross-market links provide signal beyond domestic peers.} The
Gain column in \cref{tab:best_source} compares the cross-market signal with the
same-market text baseline. It is positive for four of five
targets, showing that cross-border annual-report links add information beyond
domestic peers. Taiwan is the exception: its domestic ICIR remains higher than
any cross-market source, consistent with a tightly linked local semiconductor
cluster whose most informative peers are still at home~\citep{wang2014competitive}.

\noindent\textbf{Information flow is directional rather than symmetric.}
Reversing source and target often changes both the strength and the sign of
predictability; for example, Hong Kong predicts Taiwan, while Taiwan-to-Hong
Kong is negative. This asymmetry shows that CrossAlpha is not merely recovering
undirected business similarity, but identifies where return-relevant information
appears first.

\subsection{RQ3: Event-Conditioned Daily Spillover Evaluation with an LLM Agent}
\label{sec:rq3}
We now ask whether CrossAlpha supports a higher-frequency,
event-conditioned trading task: after a large idiosyncratic move in a source
firm, do its graph neighbours earn abnormal market-relative returns in
subsequent daily trading?

\noindent\textbf{Approach.} As shown in \cref{fig:event_pipeline}, we evaluate a post-shock
neighbour-retrieval task: after a large source-stock move, the graph retrieves
top-$K$ global neighbours, and the GPT-5 agent filters that graph shortlist into
event-specific co-movers. The agent cannot propose new firms, so the test
isolates whether agent reasoning improves graph retrieval. We use strict
$t{+}2$ open-to-open execution so all source--target time-zone orderings are
tradable. Detailed setup is in \cref{tab:rq3_protocol}. From the results in \cref{tab:event_strategy}, we have the following insights:

\begin{figure}[t]
  \centering
  \includegraphics[width=\linewidth]{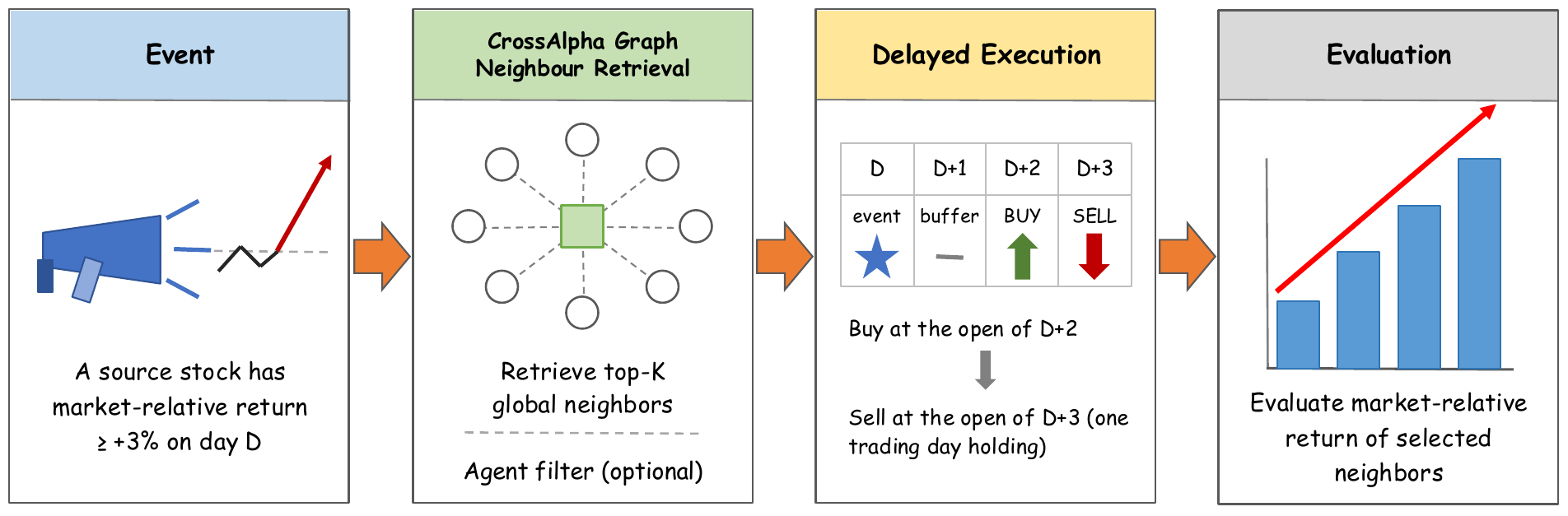}
  \caption{Event-driven evaluation pipeline.}
  \label{fig:event_pipeline}
\end{figure}

\begin{table}[t]
\centering
\caption{Event-conditioned agent evaluation setup.}
\label{tab:rq3_protocol}
\scriptsize
\setlength{\tabcolsep}{3pt}
\begin{tabular}{@{}p{0.25\columnwidth}p{0.68\columnwidth}@{}}
\toprule
\textbf{Component} & \textbf{Definition} \\
\midrule
Events & Source-stock day $D$ with market-relative close-to-close return above
+3\%; 25 source stocks fixed before the test window (top-5 by 2023-12-31 market
cap in each market; \cref{app:hub_list}). \\
Graph baskets & For each event, rank the source stock's global embedding
neighbours and evaluate equal-weight top-$K$ baskets for
$K\in\{10,20,30,60\}$. \\
Agent filter & The GPT-5 agent labels the top-60 graph neighbours as
\{co-mover, inverse, unrelated\}; \emph{Graph $+$ Agent} keeps labelled
co-movers and cannot add new firms, making the agent an auditable filter rather
than a generator. \\
Random controls & \emph{Market random} matches the graph basket's market mix;
\emph{Market+sector random} matches both market and GICS sector. Each random bar
averages 200 fixed-seed draws. \\
Timing & Strict $t{+}2$ open-to-open execution: buy at $D{+}2$ open and sell at
$D{+}3$ open. \\
Bias control & All events are after GPT-5's 2024-09-30 knowledge
cutoff\footnotemark; full construction details are in
\cref{app:event_robustness}. \\
\bottomrule
\end{tabular}
\end{table}
\footnotetext{Knowledge cutoff of OpenAI's GPT-5 \url{https://platform.openai.com/docs/models/gpt-5}.}

\begin{table}[t]
\centering
\caption{Event-conditioned cross-market transmission across basket sizes $K$
on 606 events. Graph is the CrossAlpha graph basket; Graph+Agent applies the
GPT-5 co-mover filter; the random controls match market or market+sector
composition.}
\label{tab:event_strategy}
\small 
\begin{tabular*}{\columnwidth}{@{\extracolsep{\fill}}llrrr@{}}
\toprule
\textbf{$K$} & \textbf{Basket} & \textbf{Event Ret} & \textbf{Sharpe} & \textbf{CumRet} \\
\midrule
10 & Graph & 0.107\% & 1.32 & 76.5\% \\
   & Graph+Agent & 0.157\% & 1.84 & 136.5\% \\
   & Market rand. & -0.000\% & -0.02 & -0.3\% \\
   & Market+sector rand. & -0.005\% & -0.11 & -3.3\% \\
\midrule
20 & Graph & 0.051\% & 0.81 & 30.0\% \\
   & Graph+Agent & 0.108\% & 1.50 & 80.8\% \\
   & Market rand. & -0.000\% & -0.02 & -0.2\% \\
   & Market+sector rand. & 0.008\% & 0.21 & 4.6\% \\
\midrule
30 & Graph & 0.073\% & 1.36 & 50.2\% \\
   & Graph+Agent & 0.102\% & 1.45 & 74.8\% \\
   & Market rand. & -0.003\% & -0.11 & -1.6\% \\
   & Market+sector rand. & 0.014\% & 0.44 & 8.7\% \\
\midrule
60 & Graph & 0.052\% & 1.41 & 35.1\% \\
   & Graph+Agent & 0.104\% & 1.49 & 77.3\% \\
   & Market rand. & 0.000\% & 0.01 & 0.1\% \\
   & Market+sector rand. & 0.015\% & 0.64 & 9.3\% \\
\bottomrule
\end{tabular*}
\end{table}

\noindent\textbf{CrossAlpha captures rapid and firm-specific spillovers missed by industry classifications.}  Beyond monthly factors, the disclosure-derived graph proves highly effective for daily event-driven prediction. Even under a strict $t{+}2$ execution lag—where markets have already had a full day to digest the source shock—graph neighbours generate robust abnormal returns across all basket sizes, achieving annualized Sharpe ratios up to 1.41 and cumulative returns of 76.5\%. In contrast, both market-matched and GICS-sector-matched random baskets hover near zero or suffer drawdowns. This gap proves that CrossAlpha's residual schema graph encodes precise, firm-level economic relationships that are significantly more predictive than standard industry labels.

\noindent\textbf{Agent reasoning strengthens the fixed graph shortlist.} The GPT-5 agent acts strictly as a filter without adding new names. Despite this restricted role, the agent substantially improves all return and risk-adjusted metrics across basket sizes. At $K{=}20$, the annualized Sharpe ratio rises from 0.81 to 1.50, and the cumulative compound return jumps from 30.0\% to 80.8\%. The NVIDIA case illustrates this filtering mechanism in practice. While Nintendo is structurally retrieved due to shared GPU and gaming vocabulary, the agent recognizes that the specific event concerns data-center AI rather than console demand and removes it. This reasoning proves accurate, as Nintendo subsequently realizes a -2.83\% market-relative return under the $t{+}2$ execution rule (see \cref{app:event_case_study} for details).

\section{Conclusion}
\label{sec:conclusion}
CrossAlpha is a public annual-report dataset for studying cross-market firm links through downstream return prediction. By aligning standardized disclosures, dense semantic firm graphs, and 11~years of daily prices across five equity markets, CrossAlpha moves financial-NLP evaluation beyond document-level text quality toward market-grounded firm-link reasoning. Our experiments show that disclosure-derived cross-market links capture predictive relationships missed by domestic text, industry labels, and price-only peer definitions, and that the useful source market is directed and target-specific rather than uniformly US-led. The same graph also supports event-conditioned spillover evaluation, where LLM-filtered graph neighbours improve daily post-event baskets. We release CrossAlpha as a shared benchmark bridging machine learning and quantitative finance.

\section*{Limitations}
CrossAlpha is designed as a benchmark for disclosure-grounded cross-market return prediction on liquid listed universes, not as a universal trading system. Its annual-report graph is best interpreted as a medium-horizon economic-link signal; the daily event task tests post-event spillover under a fixed feasible execution rule rather than intraday execution. Because LLMs are used to standardise filings and label event co-movers, we release the standardised text, embeddings, graphs, prompts, event lists, and evaluation outputs needed to audit the reported results.

\section*{Ethics Statement}
CrossAlpha is built from public corporate filings released through SEC EDGAR, EDINET, TWSE/MOPS, DART/KRX, and HKEX. We add no personal data, individual identities, or non-public material information, and release standardised category text rather than redistributing raw filings. All return, IC/ICIR, Sharpe, and drawdown numbers are historical academic measurements of signal quality and do not constitute investment advice, guarantees of future performance, or recommendations to trade any security. Users who deploy CrossAlpha-derived signals in live trading are responsible for their own regulatory, compliance, and risk-management obligations.

\section*{Reproducibility Statement}
The release includes the artefacts needed to reproduce the reported tables and figures: standardised ten-category text for ${\sim}10{,}700$ firm-years, PCA-whitened embeddings, the ${\sim}19$M-edge semantic graph, aligned daily OHLCV data, source-stock and event lists, event-tracing prompts, robustness controls, and JSON outputs for the reported metrics. The monthly and daily backtesting harnesses are CPU-bound and run on a single workstation. We also report the model choices and refresh cost in the appendix to make future dataset updates auditable.

\bibliography{used_citations}

\newpage
\clearpage
\onecolumn

\addcontentsline{toc}{section}{Appendix}
\part{Appendix}
\parttoc

\clearpage
\twocolumn

\appendix

\section{Data-Source Portals and Usage Policies}
\label{app:data_sources}

CrossAlpha uses public corporate-disclosure portals for annual reports and a
public market-data portal for daily OHLCV prices. \cref{tab:data_sources}
lists the source portals and the corresponding official usage-policy or terms
pages used to check redistribution and access constraints. We release
standardised category text rather than raw filings, and users refreshing the
dataset should re-check the linked policies because source terms may change.

\begin{table*}[h]
\centering
\caption{Official data-source portals and usage-policy links for CrossAlpha.}
\label{tab:data_sources}
\footnotesize
\setlength{\tabcolsep}{4pt}
\begin{tabular}{p{0.12\textwidth}p{0.20\textwidth}p{0.30\textwidth}p{0.30\textwidth}}
\toprule
\textbf{Market} & \textbf{Data stream} & \textbf{Source portal} & \textbf{Usage-policy / terms link} \\
\midrule
US & Annual reports & SEC EDGAR, \url{https://www.sec.gov/edgar} & SEC website policies and fair-access guidance, \url{https://www.sec.gov/privacy} and \url{https://www.sec.gov/os/accessing-edgar-data} \\
JP & Annual reports & EDINET, \url{https://disclosure2.edinet-fsa.go.jp} & FSA EDINET API terms, \url{https://disclosure2dl.edinet-fsa.go.jp/guide/static/disclosure/download/ESE140191.pdf} \\
TW & Annual reports & MOPS, \url{https://mops.twse.com.tw} & TWSE website terms, \url{https://wwwc.twse.com.tw/zh/terms/use.html} \\
KR & Annual reports & DART, \url{https://dart.fss.or.kr} & OpenDART terms, \url{https://opendart.fss.or.kr/intro/terms.do} \\
HK & Annual reports & HKEXnews, \url{https://www.hkexnews.hk} & HKEX website terms and conditions, \url{https://www.hkex.com.hk/Global/Exchange/Terms-of-Use} \\
All markets & Daily prices & Yahoo Finance, \url{https://finance.yahoo.com} & Yahoo terms and finance-data restrictions, \url{https://legal.yahoo.com/us/en/yahoo/terms/otos/index.html} and \url{https://legal.yahoo.com/us/en/yahoo/terms/product-atos/apiforydn/index.html} \\
\bottomrule
\end{tabular}
\end{table*}

\section{Case Studies: Cross-Market Peer Discovery}
\label{app:case_studies}

This appendix reports five examples used for manual inspection of the released cross-market graph.
The examples are drawn from the Japan TOPIX~500 universe; peer rankings are based on cosine similarity of LLM-standardized business description embeddings (\texttt{text-embedding-3-large}) computed over fiscal year 2022--2024 filings.
They are descriptive checks of the constructed links and are not used in the return-prediction evaluation.

\subsection{TSMC (2330.TW): Global Semiconductor Peers}
\label{app:tsmc}

Taiwan Semiconductor Manufacturing Company (TSMC, TSE:2330) is the world's largest dedicated semiconductor foundry.
\cref{tab:tsmc_peers} reports the top-20 cross-market peers identified by CrossAlpha.
The nearest-neighbor list is concentrated in foundry and memory firms: SMIC (SEHK:0981) ranks first, Global Foundries (NASDAQ:GFS) second, and SK~Hynix (KRX:000660) third.
SUMCO, a silicon wafer supplier, ranks fourth despite operating in an adjacent sub-segment.

\begin{table*}[h]
\centering
\caption{Top-20 cross-market peers of TSMC (2330.TW) by embedding similarity.}
\label{tab:tsmc_peers}
\begin{tabular}{rllll}
\toprule
Rank & Ticker & Company & Market & Sector \\
\midrule
1  & 0981.HK  & SMIC                          & HK & Technology \\
2  & GFS      & GlobalFoundries               & US & Technology \\
3  & 000660.KS & SK Hynix                     & KR & Technology \\
4  & 3436.T   & SUMCO                         & JP & Technology \\
5  & INTC     & Intel                         & US & Technology \\
6  & TXN      & Texas Instruments             & US & Technology \\
7  & AMAT     & Applied Materials             & US & Technology \\
8  & LRCX     & Lam Research                  & US & Technology \\
9  & KLAC     & KLA Corporation               & US & Technology \\
10 & 005930.KS & Samsung Electronics          & KR & Technology \\
11 & ASML     & ASML Holding                  & US & Technology \\
12 & MU       & Micron Technology             & US & Technology \\
13 & 6723.T   & Renesas Electronics           & JP & Technology \\
14 & NVDA     & NVIDIA                        & US & Technology \\
15 & 6954.T   & FANUC                         & JP & Technology \\
16 & ON       & ON Semiconductor              & US & Technology \\
17 & STM      & STMicroelectronics            & US & Technology \\
18 & 8035.T   & Tokyo Electron                & JP & Technology \\
19 & 009150.KS & Samsung SDI                  & KR & Technology \\
20 & ADI      & Analog Devices                & US & Technology \\
\bottomrule
\end{tabular}
\end{table*}

The market composition of the top-20 peers is: US~(12), KR~(4), JP~(3), HK~(1).
All top-10 peers fall within the Technology sector (100\% same-sector precision@10), although sector labels are not used as inputs.
The prevalence of U.S.\ names (equipment makers such as AMAT, LRCX, KLAC alongside fabless designers NVDA, TXN) reflects the tight semantic overlap between foundry operations and the semiconductor equipment and design ecosystem as described in annual reports.

\subsection{Toyota Motor (7203.T): Automotive and Cross-Sector Peers}
\label{app:toyota}

Toyota Motor Corporation (TSE:7203) is a useful example for inspecting cross-sector nearest neighbors.
\cref{tab:toyota_peers} reports the top-10 peers; the full top-30 set contains 47\% same-sector (Automotive) names alongside a set of cross-sector discoveries in technology, retail, and mobility.

\begin{table*}[h]
\centering
\caption{Top-10 cross-market peers of Toyota (7203.T) by embedding similarity.}
\label{tab:toyota_peers}
\begin{tabular}{rllll}
\toprule
Rank & Ticker & Company & Market & Sector \\
\midrule
1  & TSLA     & Tesla                 & US & Consumer Discretionary \\
2  & F        & Ford Motor            & US & Consumer Discretionary \\
3  & GM       & General Motors        & US & Consumer Discretionary \\
4  & RIVN     & Rivian Automotive     & US & Consumer Discretionary \\
5  & ORCL     & Oracle                & US & Technology \\
6  & AN       & AutoNation            & US & Consumer Discretionary \\
7  & KMX      & CarMax                & US & Consumer Discretionary \\
8  & AAPL     & Apple                 & US & Technology \\
9  & HOG      & Harley-Davidson       & US & Consumer Discretionary \\
10 & CVNA     & Carvana               & US & Consumer Discretionary \\
\bottomrule
\end{tabular}
\end{table*}

The cross-sector names at ranks 5 (Oracle) and 8 (Apple) are consistent with the overlap between automotive and technology annual reports, where vehicle manufacturers increasingly discuss software-defined platforms, in-vehicle operating systems, and data analytics infrastructure.
Further in the top-30, mobility-platform companies Uber (UBER), Lyft (LYFT), and Aurora Innovation (AUR) appear, reflecting shared business description language around autonomous driving, fleet management, and on-demand transportation.
The 47\% same-sector precision@30 indicates that the list is not simply a sector lookup.

\subsection{Apple (AAPL) $\to$ Japan: Agent-Inferred Supply Chain}
\label{app:apple_jp}

The Linker ($\phi_L$) parses 10-K filings to extract supplier and customer relationships that are not directly disclosed in structured databases.
\cref{tab:apple_jp} reports Japanese suppliers inferred for Apple from annual report disclosures, including those found in the suppliers' own filings, not solely in Apple's 10-K.

\begin{table*}[h]
\centering
\caption{Agent-inferred Japanese supply-chain peers for Apple (AAPL).}
\label{tab:apple_jp}
\begin{tabular}{lllp{6.5cm}}
\toprule
Ticker & Company & Relationship Type & Component / Role \\
\midrule
6758.T  & Sony Group        & Supplier (inferred) & Image sensors (camera modules) \\
6981.T  & Murata Mfg.       & Supplier (inferred) & Multilayer ceramic capacitors (MLCCs) \\
6762.T  & TDK               & Supplier (inferred) & Lithium-ion battery cells \\
6857.T  & Advantest         & Supplier (inferred) & Chip test and measurement equipment \\
\bottomrule
\end{tabular}
\end{table*}

None of these relationships were directly disclosed in Apple's 10-K filings; they are based on Japanese companies' own annual report disclosures referencing Apple as a major customer.
This is a common disclosure pattern: suppliers often name major customers, while customers may not list all suppliers.
The Linker therefore reads \emph{both sides} of the supply-chain relationship graph.

\subsection{NVIDIA (NVDA) $\to$ Japan: Gaming and Testing Links}
\label{app:nvidia_jp}

\begin{table*}[h]
\centering
\caption{Agent-inferred Japanese supply-chain peers for NVIDIA (NVDA).}
\label{tab:nvidia_jp}
\begin{tabular}{lllp{6.5cm}}
\toprule
Ticker & Company & Relationship Type & Component / Role \\
\midrule
6758.T  & Sony Group    & Customer (inferred) & PlayStation~5 GPU (Oberon SoC) \\
7974.T  & Nintendo      & Customer (inferred) & Nintendo Switch SoC (Tegra X1) \\
6857.T  & Advantest     & Supplier (inferred) & GPU test and verification equipment \\
\bottomrule
\end{tabular}
\end{table*}

NVIDIA's Japanese peer set combines downstream gaming console customers (Sony, Nintendo) with an upstream testing equipment supplier (Advantest).
These relationships appear because Sony and Nintendo disclose GPU vendor relationships in their product description sections, and Advantest references NVIDIA SoC platforms in its equipment disclosures.
Advantest appears in both the Apple and NVIDIA peer sets (\cref{tab:apple_jp,tab:nvidia_jp}), giving one example of a shared upstream equipment link.

\subsection{Boeing (BA) $\to$ Japan: Direct Disclosure Relationships}
\label{app:boeing_jp}

\begin{table*}[h]
\centering
\caption{Japanese supply-chain peers for Boeing (BA): direct 10-K disclosures.}
\label{tab:boeing_jp}
\begin{tabular}{lllp{6.5cm}}
\toprule
Ticker & Company & Disclosure Source & Component / Role \\
\midrule
7011.T  & Mitsubishi Heavy Industries & Boeing 10-K (direct) & Wing box structures (787 Dreamliner) \\
7012.T  & Kawasaki Heavy Industries   & Boeing 10-K (direct) & Fuselage sections, landing gear doors \\
\bottomrule
\end{tabular}
\end{table*}

Unlike the Apple and NVIDIA cases, Boeing's Japanese supplier relationships are directly named in Boeing's own 10-K filings under the supplier diversity and key suppliers disclosures.
Mitsubishi Heavy Industries is identified as the manufacturer of wing box structures for the 787 Dreamliner, and Kawasaki Heavy Industries is identified as responsible for fuselage sections and landing gear doors.
This example contrasts with the Apple and NVIDIA cases: Boeing discloses these suppliers directly, whereas some consumer-electronics links are observed from supplier-side filings.
The released graph therefore contains both directly disclosed links and links inferred from counterpart filings.

\subsection{Summary of Case Study Findings}
\label{app:case_summary}

\begin{table*}[h]
\centering
\caption{Summary statistics across the five case studies.}
\label{tab:case_summary}
\begin{tabular}{lccl}
\toprule
Company & Peer Set Size & Same-Sector \% & Key Finding \\
\midrule
TSMC (2330.TW)  & Top 20 & 100\% (top-10) & US/KR/JP foundry ecosystem \\
Toyota (7203.T) & Top 30 & 47\%            & Oracle, Apple, Uber as peers \\
Apple (AAPL)    & 4 JP links & Agent-inferred & Sony, Murata, TDK, Advantest \\
NVIDIA (NVDA)   & 3 JP links & Mixed          & Sony PS5, Nintendo Switch, Advantest \\
Boeing (BA)     & 2 JP links & Direct 10-K    & MHI wing box, KHI fuselage \\
\bottomrule
\end{tabular}
\end{table*}

The examples serve as sanity checks for three aspects of the released graph: sector recovery, cross-sector nearest-neighbor retrieval, and supplier-side disclosure links.
Evidence for return predictability is evaluated separately in the main experiments.

\section{Additional Experimental Results}
\label{app:additional_results}


\subsection{Evaluation Metric Definitions}
\label{app:metrics}
For the portfolio metrics, let $r_t^{\mathrm{LS}}$ be the net monthly long--short return after a 2~bp one-way cost and $V_t=\prod_{\tau=1}^{t}(1+r_\tau^{\mathrm{LS}})$ the corresponding wealth curve. \textbf{(1)}~\emph{IC} is the monthly rank correlation between factor values and forward returns,
\begin{equation}
\mathrm{IC}_t = \mathrm{corr}_{\mathrm{rank}}\!\left(f_i(t), r_i^{\mathrm{fwd}}(t)\right);
\label{eq:ic}
\end{equation}
\textbf{(2)}~\emph{ICIR} is its annualised information ratio,
\begin{equation}
\mathrm{ICIR} = \frac{\frac{1}{T}\sum_{t=1}^{T}\mathrm{IC}_t}{\hat{\sigma}_{\mathrm{IC}}}\sqrt{12};
\label{eq:icir}
\end{equation}
\textbf{(3)}~\emph{Sharpe} is the annualised long--short Sharpe ratio,
\begin{equation}
\mathrm{Sharpe} = \frac{\overline{r^{\mathrm{LS}}}}{\hat{\sigma}(r^{\mathrm{LS}})}\sqrt{12};
\label{eq:sharpe}
\end{equation}
\textbf{(4)}~\emph{MaxDD} is the maximum drawdown of the wealth curve,
\begin{equation}
\mathrm{MaxDD} = \min_{t}\left(\frac{V_t}{\max_{\tau\le t}V_\tau}-1\right);
\label{eq:maxdd}
\end{equation}
\textbf{(5)}~\emph{Ret} is the annualised long--short return after cost,
\begin{equation}
\mathrm{Ret} = 12\,\overline{r^{\mathrm{LS}}}.
\label{eq:ret}
\end{equation}
Finally, \textbf{(6)}~\emph{CumRet} is the ending cumulative long--short return over the evaluation window,
\begin{equation}
\mathrm{CumRet} = \sum_{t=1}^{T} r_t^{\mathrm{LS}}.
\label{eq:cumret}
\end{equation}

\subsection{Sector and Size Neutralization}
\label{app:neutralization}
For the cross-market source comparisons we neutralize each monthly cross-section in two steps. First, we demean the factor within GICS sector, removing each sector's average factor level on that date. Second, we standardize the sector-demeaned factor and regress it on standardized log market capitalisation, $\ln(\text{close}\times\text{shares outstanding})$, keeping the cross-sectional residual. The resulting factor is orthogonal to both sector membership and firm size, and the neutralized rank ICIR (\cref{eq:icir}) is computed between this residual and forward returns. A stricter variant used for robustness additionally regresses out industry- and own-momentum and winsorizes the cross-section at the $1\%$ level before standardization.

\subsection{Baseline Peer Definitions}
\label{app:baselines}
All RQ1 baselines share the monthly peer-momentum pipeline of \cref{sec:backtesting}: the same 12-month sector-relative source-market returns, the same sigmoid top-1\% weighting (\cref{eq:sigmoid}), portfolio sort, and neutralization. They differ only in how the peer weight $\alpha_{ij}$ between a target stock $i$ and a candidate $j$ is set, and---for the domestic baseline---in which market supplies the candidates.
\begin{itemize}
\item \textbf{CrossAlpha text (cross-market).} $\alpha_{ij}$ is the released US$\to$JP text similarity (cosine of PCA-whitened category embeddings, \cref{eq:text_sim}); candidates are US Russell~1000 stocks.
\item \textbf{Domestic text (JP$\to$JP).} The identical text pipeline, but candidates are the target's own market (JP TOPIX~500) and peer returns are JP sector-relative, isolating whether cross-market text adds information beyond within-market text.
\item \textbf{GICS sector.} Each target inherits the equal-weighted mean sector-relative return of the US peers sharing its GICS sector (no text and no similarity weighting); this is the industry-classification peer set.
\item \textbf{Return correlation (252d).} $\alpha_{ij}$ is the Pearson correlation of trailing 252-day daily returns between $i$ and US candidate $j$ (minimum 120 overlapping days), percentile-ranked and passed through the same sigmoid (\cref{eq:sigmoid}); this is the standard backward-looking, price-based peer definition.
\end{itemize}
Because all four aggregate peer returns identically, any performance gap reflects the peer definition alone.

\subsection{Event Strategy Robustness}
\label{app:event_robustness}

\noindent\textbf{Event execution protocol.}~Source stocks are fixed before the
test window: the top-5 firms by 2023-12-31 market capitalisation in each of the
five markets, restricted to firms with a 10-K-equivalent filing, yielding 25
stocks (\cref{app:hub_list}). On the same event set, \emph{Graph alone}
equal-weights the source stock's top-$K$ neighbours, with
$K\!\in\!\{10,20,30,60\}$, pooled across all five markets. \emph{Graph +
agent-labelled co-movers} uses the same graph ranking but keeps only candidates
that GPT-5 marks as confident co-movers before forming the $K$-name basket. The
agent never proposes firms outside the graph shortlist. Execution is strict
$t{+}2$: buy at the $D{+}2$ open and sell at the $D{+}3$ open. We leave
$D{+}1$ as a no-action buffer because different markets open in different time
zones, so a $D{+}1$ trade may be infeasible for some source--target pairs, a
standard alignment concern in cross-market return
tests~\citep{bekaert2009international,rapach2013international}. All returns are
market-relative.

\cref{tab:event_strategy} (main text) reports return and Sharpe metrics across $K$. \cref{tab:event_robustness} adds the corresponding per-event $t$-statistics on the same 606 GPT-5 post-cutoff events.

\begin{table}[h]
\centering
\caption{Per-event $t$-statistics for the event strategy in \cref{tab:event_strategy}. Same 606 GPT-5 post-cutoff events / 25 source stocks / 1.51 yrs OOS; strict $t\!+\!2$ open-to-open, market-relative returns.}
\label{tab:event_robustness}
\small
\setlength{\tabcolsep}{5pt}
\begin{tabular}{@{}lcc@{}}
\toprule
\textbf{Strategy} & \textbf{$K$} & \textbf{$t$} \\
\midrule
Graph alone (CrossAlpha emb.\ top-$K$)            & 10 & ${+}1.62$ \\
\textbf{Graph + GPT-5-labelled co-movers} & \textbf{10} & $\mathbf{{+}2.26}$ \\
\midrule
Graph alone                                       & 20 & ${+}1.00$ \\
\textbf{Graph + GPT-5-labelled co-movers} & \textbf{20} & $\mathbf{{+}1.84}$ \\
\midrule
Graph alone                                       & 30 & ${+}1.67$ \\
\textbf{Graph + GPT-5-labelled co-movers} & \textbf{30} & $\mathbf{{+}1.78}$ \\
\midrule
Graph alone                                       & 60 & ${+}1.74$ \\
\textbf{Graph + GPT-5-labelled co-movers} & \textbf{60} & $\mathbf{{+}1.83}$ \\
\bottomrule
\end{tabular}
\end{table}

\noindent\textbf{Matched-neighbour null.}~To test whether the graph result is simply a by-product of market allocation or sector composition, \cref{tab:event_neighbor_nulls} keeps the same 25 source stocks, 606 positive events, strict $t\!+\!2$ execution, and market-relative returns, but replaces each graph basket with 200 fixed-seed random baskets. The first null matches the graph basket's market allocation; the second matches each graph neighbour's market and GICS sector where available, falling back to the same market if a sector bucket is too sparse. The graph-selected neighbours sit at or above the 95th percentile of both nulls at every $K\!\in\!\{10,20,30,60\}$ (graph-percentile vs.\ market null: 100~/~99~/~100~/~100; vs.\ market$+$sector null: 100~/~95~/~99.5~/~99), indicating that the event spillover is tied to the text-derived neighbour ranking rather than to coarse market or sector exposure.

\begin{table*}[h]
\centering
\caption{Matched-neighbour null for the graph-alone event basket. Same 25 source stocks / 606 GPT-5 post-cutoff events / 1.51 yrs OOS / strict $t\!+\!2$ open-to-open / market-relative returns. Null entries report the mean per-event return across 200 fixed-seed random baskets and the 95th percentile of the seed-level null distribution.}
\label{tab:event_neighbor_nulls}
\small
\setlength{\tabcolsep}{3.5pt}
\begin{tabular}{@{}crrrrrr@{}}
\toprule
\textbf{$K$} & \textbf{Graph per-ev.} & \textbf{Graph $t$} & \textbf{Market null mean} & \textbf{Market null p95} & \textbf{Market+sector mean} & \textbf{Market+sector p95} \\
\midrule
10 & ${+}0.107\%$ & ${+}1.62$ & ${-}0.000\%$ & ${+}0.058\%$ & ${-}0.005\%$ & ${+}0.058\%$ \\
20 & ${+}0.051\%$ & ${+}1.00$ & ${-}0.000\%$ & ${+}0.040\%$ & ${+}0.008\%$ & ${+}0.051\%$ \\
30 & ${+}0.073\%$ & ${+}1.67$ & ${-}0.003\%$ & ${+}0.032\%$ & ${+}0.014\%$ & ${+}0.053\%$ \\
60 & ${+}0.052\%$ & ${+}1.74$ & ${+}0.000\%$ & ${+}0.024\%$ & ${+}0.015\%$ & ${+}0.040\%$ \\
\bottomrule
\end{tabular}
\end{table*}

\noindent\textbf{Source-stock-cluster uncertainty.}~Because events from the same source stock are not independent, we also bootstrap by resampling source stocks rather than individual events. Over 2{,}000 source-stock-cluster bootstrap draws, the graph-alone $K\!=\!30$ book has a per-event mean 95\% interval of $[-0.018\%, +0.154\%]$ and $P(\bar r>0)=0.942$. This check treats the source stock as the dependence unit and is supportive but deliberately conservative; the interval crosses zero, so we use it as a robustness diagnostic rather than as the main evidence.

\noindent\textbf{Event case study.}\label{app:event_case_study}
To make the RQ3 basket concrete, consider one post-cutoff event from the
evaluation window: SK Hynix (000660.KS) on 2025-10-02. The source stock had a
market-relative return of ${+}9.11\%$, consistent with a positive AI-memory /
HBM demand shock. CrossAlpha's graph retrieved a global neighbour shortlist
containing memory-sector peers and hardware customers; the agent then labelled
confident positive co-movers such as Taiwanese semiconductor peers
2408.TW, 6531.TW, 8271.TW, and 2344.TW, plus hardware-exposed 4967.TW. Under the
same strict $t{+}2$ rule as the main experiment, the confident co-mover basket
with available cross-market returns earned an average market-relative return of
${+}4.53\%$ over the one-day $D{+}2$ open-to-$D{+}3$ open holding window, with
15 of 17 names positive. This example is descriptive rather than a separate
test: the reported RQ3 results pool all 606 post-cutoff events.

The agent also removes graph neighbours that are textually close but have the
wrong exposure for the event. For an NVIDIA AI-demand shock, the graph shortlist
contains Nintendo (7974.T), a plausible textual neighbour because both firms
share gaming and GPU-related business language: NVIDIA historically sells GPUs
into gaming devices, while Nintendo discloses console hardware, game software,
and platform ecosystems. The textual graph therefore retrieves Nintendo as a
reasonable broad business neighbour. The agent labels Nintendo as
\emph{unrelated} with sign~0, because the event is a data-center AI GPU demand
surprise rather than a console-cycle or gaming-demand shock. In that event
window, firms tied to AI accelerators, HBM memory, servers, and semiconductor
equipment should be more direct co-movers, while Nintendo's economics are
driven mainly by Switch hardware, first-party software, and consumer gaming
demand. The realised return is consistent with this filter: under the same
strict $t{+}2$ open-to-open rule after NVIDIA's 2024-02-22 AI-demand event,
Nintendo's one-day market-relative return was $-2.83\%$ (raw return
$-2.65\%$), so it did not participate in the positive AI-semiconductor
spillover. This is the intended role of the agent filter: keep firms with
direct exposure to the shock and remove semantically nearby firms whose
economics should not co-move in that event window.

\noindent\textbf{Source-stock list.}\label{app:hub_list} The 25 source stocks used in RQ3, by region and Dec-2023 mcap rank: \textbf{US Russell~1000}: AAPL, MSFT, GOOG, AMZN, NVDA; \textbf{JP TOPIX~500}: 7203.T (Toyota), 8306.T (Mitsubishi UFJ), 6758.T (Sony), 6861.T (Keyence), 8035.T (Tokyo Electron); \textbf{TW TWSE}: 2330.TW (TSMC), 2317.TW (Hon Hai / Foxconn), 2454.TW (MediaTek), 2308.TW (Delta Electronics), 2382.TW (Quanta); \textbf{KR KOSPI}: 005930.KS (Samsung Electronics), 000660.KS (SK Hynix), 373220.KS (LG Energy Solution), 207940.KS (Samsung Biologics), 005380.KS (Hyundai Motor); \textbf{HK Main}: 0700.HK (Tencent), 9988.HK (Alibaba), 1398.HK (ICBC), 0939.HK (CCB), 0005.HK (HSBC).

\subsection{LLM Extraction Quality}
\label{app:extraction_quality}

We assess LLM extraction quality via a human-annotated sample of 200 randomly selected filings (25 per market), with three categories randomly chosen per filing.
Annotators rate accuracy on a 3-point scale: ``accurate'' (89\%), ``partial'' (9\%), and ``incorrect'' (2\%).
The lowest accuracy is for the Korean-language filings (82\%), consistent with lower LLM pretraining coverage for non-English financial disclosures.

Bilingual Hong Kong filings provide a consistency check: the average cosine similarity between English and Chinese sections of the same filing is 0.84 ($\pm$0.06), suggesting that the Distiller produces stable business descriptions when the same issuer discloses in both versions.

\subsection{Dense Graph Construction Details}
\label{app:graph_details}

This appendix gives the construction detail condensed in \cref{sec:graph_construction}, where the directed cross-market text-similarity graph $\mathcal{G}_{\mathrm{text}}=(V,E_{\mathrm{text}})$ and the firm--firm similarity $s_{\mathrm{text}}(i,j)$ are defined (\cref{eq:text_graph,eq:text_sim}). Each node additionally carries a home-market type via $\tau(v): V \to \mathcal{M}$. We use equal category weights in $s_{\mathrm{text}}$ to avoid injecting task-specific beliefs about which disclosure fields should matter most; category-specific and learned-weight variants are treated as ablations rather than the default construction.

\noindent \textbf{Peer-selection weights.}
To form discrete edges $(i,j) \in E_{\mathrm{text}}$ and concentrate signal on the most relevant peers, we apply percentile ranking and a sigmoid transformation:
\begin{equation}
\begin{aligned}
\alpha_{ij}^{\mathrm{text}}
&= \left[1 + \exp\!\left(-\kappa(\mathrm{rank}_{ij} - \tau)\right)\right]^{-1},\\
\kappa &= 50,\qquad \tau = 0.99 .
\end{aligned}
\label{eq:sigmoid}
\end{equation}
This assigns near-zero weight to the majority and concentrates $\alpha_{ij}^{\mathrm{text}}$ on the top ${\sim}$1\% most similar peers, roughly 10 stocks per target.

\noindent \textbf{Embedding and whitening.}
Each standardised category text is embedded using OpenAI \texttt{text-embedding-3-large}~\citep{openai2024embedding}, yielding
\begin{equation}
\mathbf{e}_i^{(c)} \in \mathbb{R}^{3072}.
\label{eq:raw_embedding}
\end{equation}
Text embeddings are well known to suffer from a long-tailed eigen-spectrum and semantic collapse, where a few dominant directions inflate pairwise cosine similarities~\citep{su2021whitening}; here these directions capture the shared style of the standardised summaries rather than firm-specific content. We address this with PCA \citep{kessy2018optimal}, projecting embeddings onto an isotropic, lower-dimensional space:
\begin{equation}
\mathbf{z}_i^{(c)} = \sqrt{n} \cdot \mathbf{U}_d \boldsymbol{\Sigma}_d^{-1} \mathbf{e}_i^{(c)},
\label{eq:whitening}
\end{equation}
where $\mathbf{z}_i^{(c)} \in \mathbb{R}^d$, and $\mathbf{U}_d$ and $\boldsymbol{\Sigma}_d$ are the top-$d$ components from the SVD of the centred embedding matrix. CrossAlpha releases six dimensionality variants,
\begin{equation}
d \in \{64, 128, 256, 512, 768, 1024\},
\label{eq:pca_dims}
\end{equation}
so users can revisit the trade-off on new markets.

\subsection{Hyperparameter Sensitivity}
\label{sec:ablations}

\begin{figure}[t]
\centering
\begin{subfigure}[t]{\linewidth}
\centering
\begin{tikzpicture}
\begin{axis}[
  width=0.70\linewidth, height=3.15cm,
  xmode=log, log basis x=2,
  xtick={64,128,256,512,1024},
  xticklabels={64,128,256,512,1024},
  xlabel={\scriptsize PCA dimension $d$},
  ylabel={\scriptsize raw ICIR},
  xlabel style={yshift=3pt}, ylabel style={yshift=-4pt},
  ymin=0.30, ymax=0.55,
  ytick={0.30,0.35,0.40,0.45,0.50,0.55},
  axis line style={draw=gray!60}, tick style={draw=gray!60},
  x tick label style={font=\tiny}, y tick label style={font=\tiny},
  xmajorgrids, ymajorgrids, grid style={gray!20, dashed}
]
\addplot[uscolor, line width=0.9pt, mark=*, mark size=1.6pt, mark options={fill=uscolor, draw=uscolor}] coordinates {
(64,0.42) (128,0.50) (256,0.48) (512,0.44) (768,0.40) (1024,0.37)
};
\node[font=\tiny\bfseries, uscolor!80!black] at (axis cs:128, 0.535) {peak};
\end{axis}
\end{tikzpicture}
\caption{Embedding size $d$.}
\label{fig:abl_dim}
\end{subfigure}
\vspace{0.4em}

\begin{subfigure}[t]{\linewidth}
\centering
\begin{tikzpicture}
\begin{axis}[
  width=0.70\linewidth, height=3.15cm,
  xlabel={\scriptsize Lookback $L$ (months)},
  ylabel={\scriptsize Monthly rank ICIR},
  xlabel style={yshift=3pt}, ylabel style={yshift=-4pt},
  ymin=0.10, ymax=0.50,
  xtick={1,3,6,12,24,36},
  ytick={0.1,0.2,0.3,0.4,0.5},
  axis line style={draw=gray!60}, tick style={draw=gray!60},
  x tick label style={font=\tiny}, y tick label style={font=\tiny},
  xmajorgrids, ymajorgrids, grid style={gray!20, dashed}
]
\addplot[jpcolor!85!black, line width=0.9pt, mark=square*, mark size=1.7pt, mark options={fill=jpcolor, draw=jpcolor!80!black}] coordinates {
(1,0.22) (3,0.17) (6,0.36) (12,0.43) (24,0.35) (36,0.26)
};
\node[font=\tiny\bfseries, jpcolor!80!black] at (axis cs:12, 0.48) {peak};
\end{axis}
\end{tikzpicture}
\caption{Lookback $L$.}
\label{fig:abl_lookback}
\end{subfigure}
\caption{Hyperparameter sensitivity of the US-to-JP factor. (a)~Raw ICIR peaks at PCA dimension 128 and declines beyond. (b)~Monthly rank ICIR peaks at a 12-month lookback and decays on both sides.}
\label{fig:ablations}
\end{figure}
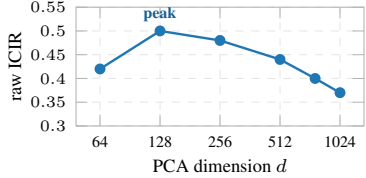
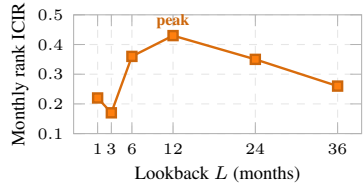

We test how sensitive the cross-market factor is to the two settings that govern the dense-graph pipeline's inductive bias, the PCA-whitening dimension and the peer-return lookback, sweeping each while holding everything else at the defaults used throughout the paper (\cref{fig:ablations}).

\noindent\textbf{Sensitivity around the released defaults.} In this sweep, the released defaults are near the best observed settings. Rank ICIR peaks at the PCA-whitening dimension of 128 (0.50) and declines monotonically to 0.37 at 1024, consistent with embedding anisotropy and noisy tail components~\citep{su2021whitening}. The peer-return lookback peaks at 12 months (0.43) and decays on both sides, consistent with the annual-filing cycle and the customer--supplier diffusion timescale~\citep{cohen2008economic,menzly2010market}. The preferred dimension is mildly market-dependent (e.g., US$\to$TW semiconductors favour $d{=}256$); all six variants ($d \in \{64,128,256,512,768,1024\}$) are released so researchers can test this sensitivity.

\subsection{Embedding-Model Provider Robustness}
\label{app:embedding_provider_robustness}

To check whether the text-similarity signal depends on one embedding provider, we run a lightweight robustness check on the US$\to$Japan source--target setting using only the \texttt{main\_business\_segments} field (\cref{tab:embedding_provider_robustness}). This diagnostic uses the daily peer-momentum harness rather than the headline monthly neutralized pipeline, so we treat it as a representation sanity check rather than a replacement for \cref{tab:baselines_monthly}. We compare OpenAI \texttt{text-embedding-3-large} against two open-source sentence-transformer families (MiniLM/SBERT~\citep{wang2020minilm,reimers2019sentencebert} and MPNet/SBERT~\citep{song2020mpnet,reimers2019sentencebert}), plus FinBERT~\citep{araci2019finbert} and a sparse TF-IDF baseline~\citep{salton1975vector}. All neural and sparse representations produce positive ICIR under the same US-peer / JP-target factor construction; OpenAI remains the strongest single encoder in this diagnostic, while MPNet and TF-IDF are close.

\begin{table}[h]
\centering
\caption{Embedding-provider robustness on US$\to$Japan using \texttt{main\_business\_segments} only. ICIR is daily rank-ICIR under the same peer-momentum harness at 5-day and 20-day peer-return lookbacks.}
\label{tab:embedding_provider_robustness}
\scriptsize
\setlength{\tabcolsep}{3pt}
\begin{tabular*}{\linewidth}{@{\extracolsep{\fill}}lrrr@{}}
\toprule
\textbf{Representation} & \textbf{Dim.} & \textbf{5d ICIR} & \textbf{20d ICIR} \\
\midrule
OpenAI 3-large & 3072 & $6.33$ & $2.79$ \\
MPNet/SBERT    & 768  & $6.04$ & $2.73$ \\
TF-IDF         & 5000 & $6.04$ & $2.74$ \\
MiniLM/SBERT   & 384  & $5.76$ & $2.58$ \\
FinBERT        & 768  & $5.38$ & $2.33$ \\
\bottomrule
\end{tabular*}
\end{table}

\section{More Related Work}
\label{app:related_detail}

This appendix expands the main-paper related-work summary (\cref{sec:related_work}) with additional detail on the research areas CrossAlpha draws on.

\noindent \textbf{Cross-market Return Predictability.} Classic international-finance work documents return and volatility transmission across equity markets~\citep{eun1989international,hamao1990correlations,king1990transmission}, and later studies show that US returns, international co-movement, and trade links help explain global market linkages~\citep{rapach2013international,bekaert2009international,forbes2004decomposition}. At the firm level, customer--supplier, same-industry, and production-network links generate predictable return spillovers~\citep{cohen2008economic,menzly2010market,Acemoglu2012,herskovic2018networks}. Recent cross-market forecasting work mainly learns from price/volume series or trading-signal graphs~\citep{cspo2025,interintragnn2025,liu2026bipartite}, while text-derived firm-link studies either remain mostly domestic or are not released as cross-market return benchmarks~\citep{feng2019temporal,kim2019hats,hybridfirm2025,breitung2025global}. CrossAlpha fills this gap by releasing annual-report schema text, cross-market semantic links, aligned prices, and factor-evaluation protocols.

\noindent \textbf{Text-based Firm Similarity.} Text-based network industries derive peer links from 10-K product descriptions~\citep{hoberg2010product,hoberg2016text}. More recent work uses contextual embeddings to discover economic links for momentum spillover, while related stock-embedding methods combine text with network information~\citep{chung2023modeling,setn2024}. Topic-based studies similarly show that peer returns extracted from MD\&A narratives or earnings-call business aspects predict focal-firm returns~\citep{zhang2026uncovering,jin2024business}. These pipelines build on financial text representations, from sentiment lexicons and domain language models~\citep{tetlock2007giving,loughran2011liability,araci2019finbert,wu2023bloomberggpt} to modern embedding backbones~\citep{reimers2019sentencebert,xiao2023bgeembedding}. Financial NLP benchmarks cover language understanding and embedding retrieval or similarity~\citep{shah2022flue,jorgensen2023multifin,finsts2024,tang2025finmteb}. However, these resources score linguistic or document-level quality rather than firm-level return prediction, and most text-derived peer-return studies remain confined to a single market.

\subsection{Financial NLP Benchmarks}

Financial NLP benchmarks standardise evaluation for sentiment, classification, retrieval, and semantic-similarity tasks on financial text, but none of them ship the firm-level cross-market return-prediction surface CrossAlpha is built around.
\citet{araci2019finbert} introduces FinBERT, a domain-adapted language model for financial sentiment.
\citet{shah2022flue} releases FLUE, a multi-task benchmark spanning English-language financial text classification.
\citet{jorgensen2023multifin} extends benchmark coverage beyond English financial documents but at the document-classification level rather than firm-level peer linking.
\citet{finsts2024} introduce a financial narrative semantic-similarity task for detecting subtle statement-level shifts.
\citet{tang2025finmteb} releases FinMTEB, a massive embedding benchmark covering retrieval and similarity but not return prediction.
These resources are valuable model-evaluation suites, but their supervision is primarily linguistic or document-level.
They do not ship a firm-level panel that aligns annual reports, daily prices, cross-market peer links, and downstream factor-construction code.
A representation that performs well on these NLP tasks may still fail the economic test that matters for asset pricing: selecting firms whose future returns are predictable from economically related firms' disclosures.

\subsection{Cross-Market and Graph-Based Return Models}

Classic international-finance studies establish that shocks, returns, and volatility are transmitted across major equity markets~\citep{eun1989international,hamao1990correlations,king1990transmission}.
Related decompositions show that cross-country stock-market linkages reflect both global financial factors and real economic channels such as trade~\citep{forbes2004decomposition}.
A more recent line of work moves beyond a single exchange by training prediction models on market time-series features across multiple markets, or by augmenting return prediction with graph structure.
\citet{cspo2025} train a cross-market predictor over market time-series features and show it outperforms single-market baselines.
\citet{interintragnn2025} use a graph neural network with intra- and inter-market message passing for cross-market return forecasting.
\citet{liu2026bipartite} build a U.S.--China bipartite graph for cross-market return forecasting.
\citet{feng2019temporal} and \citet{kim2019hats} are representative domestic relational stock-prediction models: they use company relations or relation-type attention to improve stock ranking and movement prediction.
\citet{hybridfirm2025} augment a return-prediction architecture with a sparse supply-chain or money-flow graph alongside firm-level features.
These methods are complementary to CrossAlpha but do not solve the dataset problem.
Price-only cross-market models observe co-movement after the fact but do not explain why two firms should be economically linked, while existing graph datasets either remain domestic or are not released as reusable cross-border annual-report resources.
CrossAlpha instead provides text-grounded firm similarity across five markets and a typed, directed US$\to$Japan economic-linkage overlay, enabling both predictive evaluation and relationship-level interpretation.

\subsection{Text-Derived Firm Networks and Peer Momentum}

Text-derived peer networks are the closest methodological neighbour to CrossAlpha.
\citet{hoberg2016text} construct text-based network industries from 10-K product descriptions, establishing that disclosure text can recover economically meaningful peer sets.
\citet{chung2023modeling} use financial documents to discover economic links for momentum spillover, and \citet{setn2024} combine textual and network information for stock embeddings.
\citet{zhang2026uncovering} construct MD\&A topic peers and show that peer lagged returns predict focal-firm returns in the Chinese market.
\citet{jin2024business} reaches a similar conclusion from earnings conference-call topic similarity, where business-aspect peers reveal difficult-to-observe firm relatedness.
Most directly related, \citet{breitung2025global} use LLM-generated historical business descriptions and OpenAI embeddings to construct global business networks and study global-stock lead--lag effects.
CrossAlpha builds on this direction but changes the unit of contribution: it releases a multi-market annual-report benchmark with standardized schema fields, dense cross-market firm links, aligned OHLCV data, and reproducible factor protocols, rather than only demonstrating that text-derived business networks are useful in downstream finance applications.

\subsection{LLMs and Agents for Financial Extraction}

General-purpose LLMs have been used to extract forecast-useful information from financial news and filings.
\citet{lopezlira2023chatgpt} show that GPT sentiment scores on news headlines predict next-day equity returns in the US.
\citet{kim2024financial} demonstrate that GPT-4 with chain-of-thought reasoning over earnings releases surpasses human analysts in binary earnings-direction forecasting.
On the agentic side, \citet{xiao2024tradingagents} use a multi-step LLM workflow that combines news, filings, and price context to issue trading decisions.
\citet{fatouros2025marketsenseai} construct MarketSenseAI, an agent system that explicitly retrieves and reasons over corporate disclosures for market analysis.
CrossAlpha uses LLMs differently: not as a trading agent, but as dataset-construction infrastructure.
The Distiller standardises heterogeneous annual reports into a shared business schema, and the Linker mines typed, directed economic relationships that are difficult to recover from surface mentions alone.
This stores the LLM outputs as reusable data fields that can be evaluated with conventional asset-pricing metrics.

\subsection{Supply-Chain Networks in Asset Pricing}

\citet{Acemoglu2012} show that idiosyncratic firm-level shocks aggregate into sizeable macroeconomic fluctuations when the production network is sufficiently asymmetric.
\citet{herskovic2018networks} embed production networks into an asset-pricing model and show that network-implied exposure generates cross-sectional return predictability: firms with similar upstream and downstream connectivity earn correlated excess returns.
Empirical identification of supply-chain effects is challenging because supply linkages are typically proprietary; \citet{barrot2016input} exploit natural disasters as quasi-experiments, showing that input specificity amplifies shock propagation, with downstream equity prices falling significantly following upstream disruptions.
These findings motivate CrossAlpha's Linker pass: by parsing relationship disclosures from US 10-K filings (and cross-reading the Asian counterparties' own filings), we operationalise the production-network channel at scale, achieving a 47$\times$ coverage expansion over direct mentions.

\section{Release, Reproducibility, and Construction Cost}
\label{app:release}

\subsection{The \texttt{AR-Scraper} Toolkit}
\label{app:ar_scraper}
To make CrossAlpha a \emph{living} benchmark rather than a one-shot dump, the anonymized release includes \texttt{AR-Scraper}, a modular Python toolkit that exposes one scraper module per regulatory system behind a common \texttt{fetch(ticker, fy)} interface (\cref{tab:ar_scraper}), together with the Distiller prompt used in this work. Researchers can therefore reproduce the full pipeline end-to-end, from raw filings to evaluated factors, on new vintages, new markets, or alternative LLM backends. All data will be released under CC-BY-4.0 and code under MIT.

\begin{table}[h]
\centering
\caption{The five \texttt{AR-Scraper} modules shipped with CrossAlpha. Module names are suffixes under the \texttt{ar\_scraper} package; each implements the same \texttt{fetch(ticker, fy)} interface.}
\label{tab:ar_scraper}
\scriptsize
\setlength{\tabcolsep}{2pt}
\begin{tabular*}{\linewidth}{@{\extracolsep{\fill}}lllcc@{}}
\toprule
\textbf{Module} & \textbf{Market} & \textbf{System} & \textbf{Lang.} & \textbf{Doc type} \\
\midrule
\texttt{edgar}  & US & SEC EDGAR & en             & 10-K \\
\texttt{edinet} & JP & EDINET    & ja             & \emph{yuho} \\
\texttt{mops}   & TW & MOPS      & zh-Hant        & annual report \\
\texttt{dart}   & KR & DART      & ko             & business report \\
\texttt{hkex}   & HK & HKEx      & en\,/\,zh-Hant & annual report \\
\bottomrule
\end{tabular*}
\end{table}

\subsection{Construction Cost and Latency}
\label{app:cost}
\noindent \textbf{LLM spend per dataset refresh.}
All figures are computed from public pricing tiers as of early 2026.
(1)~\emph{Distiller pass} uses GPT-4.1 at \$2 / \$8 per million input / output tokens, averaging ${\approx}$40k input and ${\approx}$2k output tokens per filing, for ${\approx}$\$0.10 per filing or \textbf{${\approx}$\$1{,}000 total} across ${\approx}$10{,}700 filings.
(2)~\emph{Dense embedding} uses OpenAI \texttt{text-embedding-3-large} at \$0.13/M tokens; ten category strings per filing at ${\approx}$2k tokens each cost \textbf{${\approx}$\$5 total}.
(3)~\emph{Linker pass} (US$\to$JP typed-edge layer) uses GPT-4.1 on ${\approx}$9{,}200 candidate pairs at ${\approx}$5k input and ${\approx}$0.5k output tokens per call, for \textbf{${\approx}$\$25 total}; the open-universe retriever variant adds another \textbf{${\approx}$\$30}.
(4)~The \emph{event-tracing} agent (\cref{sec:rq3}) uses a single GPT-5 call per source stock on ${\approx}$60 neighbours (graph$+$typing book) or on the full $\sim$2{,}000-firm universe (agent-solo ablation) for 25 source stocks at a fixed cost of \textbf{${\approx}$\$8 total}; the typed output is the per-source-stock annotation we release as part of the event-tracing benchmark.
Together, a full CrossAlpha refresh costs about \textbf{\$1{,}070} in API spend (Distiller \$1{,}000 + embedding \$5 + Linker \$25 + retriever \$30 + agent \$8, rounded), a one-time inference budget for the entire cross-market benchmark.

\noindent \textbf{Wall-clock latency.}
Respecting provider rate limits with $\le 20$ concurrent requests, the Distiller pass completes in ${\approx}$8~hours and embedding in ${<}$1~hour. Backtesting the monthly factor harness is CPU-bound and completes in ${\approx}$30~minutes on a single workstation. The base dense benchmark can therefore be refreshed within a working day.

\section{Distiller and Linker Prompts}
\label{app:prompts}

This appendix gives the prompt templates used to instantiate the two role-specialised LLM passes introduced in \cref{sec:data_collection}: the \textbf{Distiller} ($\phi_D$, GPT-4.1) for market-level filing standardisation and the \textbf{Linker} ($\phi_L$, GPT-4.1, same backbone, relational-reasoning role) for typed-edge mining. Both prompts use temperature~0 and require strict JSON output; field names match those used in tables and figures throughout the main text. The full reference implementation, including parsing helpers, retry logic, and per-market wrappers, ships with the \texttt{AR-Scraper} toolkit (\cref{app:ar_scraper}).

\subsection{Distiller Prompt ($\phi_D$): 10-Category Standardisation}
\label{app:distiller_prompt}

Each filing is processed independently. The system message is shared across all five markets; only the user message changes per filing. Output tokens are capped at 2{,}000 to control cost; on truncation we retry the offending field individually.

\begin{tcolorbox}[colback=gray!5, colframe=blue!35!black, colbacktitle=blue!8, coltitle=black, title={Distiller system message ($\phi_D$, GPT-4.1)}, fonttitle=\bfseries\small, fontupper=\small]
\raggedright
You are an expert multilingual financial analyst fluent in English, Japanese, Chinese (Traditional and Simplified), and Korean. Your task is to read a corporate annual report (10-K, yuho, business report, or equivalent) and distil it into ten structured English business-description fields.\\[2pt]
Rules:\\
1. Output English only, even when the source filing is in another language.\\
2. Be concise: each field $\le$ 80 tokens.\\
3. Use only facts present in or directly entailed by the filing; do not import outside knowledge.\\
4. Avoid boilerplate (risk-factor language, forward-looking caveats, generic mission statements).\\
5. If a field cannot be inferred from the filing, output \texttt{null}.\\
6. Respond ONLY with a single JSON object; no explanatory text, no markdown.
\end{tcolorbox}

\begin{tcolorbox}[colback=gray!5, colframe=blue!35!black, colbacktitle=blue!8, coltitle=black, title={Distiller user message template ($\phi_D$)}, fonttitle=\bfseries\small, fontupper=\small]
\raggedright
TICKER: \{\{ticker\}\}\\
MARKET: \{\{market\}\}\\
FISCAL\_YEAR: \{\{fy\}\}\\
FILING:\\
\{\{filing\_text\}\}\\[4pt]
Return a JSON object with EXACTLY these ten keys (string values, English):\\
\quad 1. \texttt{main\_business\_segments}\\
\quad 2. \texttt{core\_technologies}\\
\quad 3. \texttt{primary\_customers}\\
\quad 4. \texttt{supply\_chain\_position}\\
\quad 5. \texttt{geographic\_coverage}\\
\quad 6. \texttt{financial\_profile}\\
\quad 7. \texttt{revenue\_model}\\
\quad 8. \texttt{value\_proposition}\\
\quad 9. \texttt{strategic\_focus}\\
\quad 10. \texttt{key\_competitors}
\end{tcolorbox}

\subsection{Linker Prompt ($\phi_L$): Typed Cross-Market Edge Mining}
\label{app:linker_prompt}

For the US$\to$JP source--target setting, the Linker is invoked once per US 10-K with the candidate JP universe (TOPIX 500 tickers + GICS sectors) injected into the user message. The dual-side reading mechanism is encoded by allowing the agent to mark an edge as supported by \texttt{supplier\_side\_disclosure}, i.e., the edge is recovered from the \emph{target's} \emph{yuho} rather than the source 10-K. For instance, Apple's 10-K names no Japanese firm (only generic ``outsourcing partners in Asia''), yet reading the targets' \emph{yuho} yields typed supplier edges from Apple to Sony, Murata, TDK, and Advantest; these links are not explicit in Apple's own filing.

\begin{tcolorbox}[colback=gray!5, colframe=blue!35!black, colbacktitle=blue!8, coltitle=black, title={Linker system message ($\phi_L$, GPT-4.1)}, fonttitle=\bfseries\small, fontupper=\small]
\raggedright
You are an expert in global supply chains, corporate disclosure, and cross-border equity research. You will be given (a) a US 10-K filing and (b) a target foreign market. Your task is to enumerate every named or strongly inferable foreign firm in the target market that has one of four typed relationships to the source firm:\\[2pt]
\quad \texttt{SUPPLIER}\quad: provides inputs, components, or services to the source.\\
\quad \texttt{CUSTOMER}\quad: purchases material volume from the source.\\
\quad \texttt{PARTNER}\quad: joint venture, alliance, or co-development.\\
\quad \texttt{COMPETITOR}: directly overlapping product or service market.\\[2pt]
Rules:\\
1. Combine explicit text mentions in the source 10-K with grounded world knowledge (e.g., a documented major customer relationship reported in the supplier's own filing).\\
2. When the 10-K is silent but the relationship is well-attested in the target firm's public disclosures, mark \texttt{evidence = "supplier\_side\_disclosure"} and lower the confidence accordingly.\\
3. Do NOT include US-domestic firms; relations must cross the source/target border.\\
4. Provide a one-sentence \texttt{rationale} naming the specific product, segment, or contract that grounds the edge.\\
5. Respond ONLY with strict JSON.
\end{tcolorbox}

\begin{tcolorbox}[colback=gray!5, colframe=blue!35!black, colbacktitle=blue!8, coltitle=black, title={Linker user message template ($\phi_L$)}, fonttitle=\bfseries\small, fontupper=\small]
\raggedright
SOURCE\_TICKER: \{\{us\_ticker\}\}\\
SOURCE\_FILING (10-K):\\
\{\{us\_10k\_text\}\}\\[4pt]
TARGET\_MARKET: \{\{market\}\}\\
TARGET\_UNIVERSE (ticker, GICS sector):\\
\{\{candidate\_table\}\}\\[4pt]
Return a JSON list. Each element:\\
\{\\
\quad "target\_ticker":\ string,\\
\quad "relation\_type":\ "supplier" | "customer" | "partner" | "competitor",\\
\quad "evidence":\ "explicit\_mention" | "inferred\_world\_knowledge" | "supplier\_side\_disclosure",\\
\quad "confidence":\ float in [0, 1],\\
\quad "rationale":\ string ($\le$ 40 words)\\
\}
\end{tcolorbox}

\noindent \textbf{Static-edge-mining variants.}
The Linker variants that produced the typed-edge release reuse the system message above with the following changes to the user message:
$\phi_L^{\mathrm{A}}$ (per-pair grader) is invoked once per (US, JP) candidate pair from the embedding top-20 and outputs a single edge with a continuous strength score $s \in [0,1]$;
$\phi_L^{\mathrm{B}}$ (comparative re-ranker) receives all $K{=}50$ embedding candidates in one prompt and returns a ranked top-10 with per-pair strengths;
$\phi_L^{\mathrm{C}}$ (retriever) drops the candidate table and instead exposes a \texttt{search(query)} tool over the full Russell~1000.

\subsection{Event-Tracing Prompt ($\phi_L^{\mathrm{event}}$): Typed Sign on a Hub Event}
\label{app:event_tracing_prompt}

The event-tracing prompt of \cref{sec:rq3} is invoked \emph{once per source stock} (not per pair), takes as input the list of the source stock's top-$K$ dense-graph cross-market neighbours with their identity fields, and returns a typed sign + confidence + relation for each. The system message instantiates the source-stock-specific knowledge in-place (here SK~Hynix; one would swap in the source stock's identity for a different test).

\begin{tcolorbox}[colback=gray!5, colframe=blue!35!black, colbacktitle=blue!8, coltitle=black, title={Event-tracing system message ($\phi_L^{\mathrm{event}}$, GPT-4.1, temperature 0)}, fonttitle=\bfseries\small, fontupper=\small]
\raggedright
You are a sell-side equity analyst who reasons about cross-market spillover. SK Hynix (000660.KS, South Korea, semiconductor / DRAM \& HBM memory manufacturing) has just had a LARGE POSITIVE IDIOSYNCRATIC stock move (a memory-demand or memory-pricing surprise, the typical SK Hynix shock in the 2023-26 AI cycle). For each firm in the list, predict the sign of its ABNORMAL (market-relative) stock return over the NEXT 1-3 trading days, reasoning about the economic link:\\[2pt]
\quad sign = +1 -- co-mover: semiconductor-equipment / materials supplier to memory makers; memory or broad-semis sector peer; AI-accelerator or HBM customer whose demand is validated; fab / foundry whose capex outlook improves.\\
\quad sign = -1 -- inverse mover: a firm whose MAJOR INPUT COST is DRAM/NAND/HBM memory and has little offsetting upside (e.g.\ a pure PC / server / handset / consumer-electronics OEM that BUYS memory).\\
\quad sign =  0 -- no material economic exposure to a SK Hynix memory shock.\\[2pt]
Respond ONLY with a JSON list, one object per firm, in the SAME ORDER as the input:\\
\{"ticker":\ "...",\\
\ "relation":\ "<supplier|customer|sector\_peer|\\
\quad\ foundry|memory\_consumer|unrelated|...>",\\
\ "sign":\ -1|0|1,\ "confidence":\ <float 0-1>,\\
\ "rationale":\ "<<=25 words>"\}
\end{tcolorbox}

\begin{tcolorbox}[colback=gray!5, colframe=blue!35!black, colbacktitle=blue!8, coltitle=black, title={Event-tracing user message template ($\phi_L^{\mathrm{event}}$)}, fonttitle=\bfseries\small, fontupper=\small]
\raggedright
FIRMS (SK Hynix's CrossAlpha embedding neighbours, in similarity-rank order):\\
- \{\{ticker\}\} | \{\{name\}\} | \{\{country\}\} | industry: \{\{industry\}\} | GICS: \{\{gics\}\} | (market: \{\{region\}\})\\
- ... (one line per neighbour, global top-$K\!=\!60$ pooled across the three non-home markets)
\end{tcolorbox}

The benchmark protocol scores any future agent by running its labelled-sign output through the same long-only-confident-${+}$ backtest of \cref{tab:event_robustness} on the same 25-source-stock / 606-event GPT-5 post-cutoff window, so improvements are measured in downstream predictive performance, not in label-matching against a noisy gold set.

\section{Grounding the 10-Category Schema Across Regulatory Systems}
\label{app:schema_grounding}

The ten Distiller categories (main\_business\_segments, core\_technologies, primary\_customers, supply\_chain\_position, geographic\_coverage, financial\_profile, revenue\_model, value\_proposition, strategic\_focus, key\_competitors) are chosen to satisfy two constraints. First, each
category corresponds to a cross-firm channel that prior firm-level
return-predictability research finds informative (\cref{tab:schema_rationale}).
Second, every category is recoverable from \emph{every} mandatory annual filing
across the five regulatory systems we cover (\cref{tab:schema_grounding}).
This cross-system grounding is what makes the Distiller's schema a true
lowest-common-denominator: no category depends on disclosure conventions unique
to a single market.

\begin{table*}[h]
\centering
\caption{\textbf{The ten Distiller categories and why each is included.} Each category captures a cross-firm channel that firm-level return-predictability research finds informative; the full set is the intersection of these channels with content mandated across all five filing systems.}
\label{tab:schema_rationale}
\small
\setlength{\tabcolsep}{6pt}
\renewcommand{\arraystretch}{1.15}
\begin{tabular}{@{}llp{0.3\linewidth}@{}}
\toprule
\textbf{Category} & \textbf{Cross-firm channel it captures} & \textbf{Motivated by} \\
\midrule
\texttt{main\_business\_segments} & Product-market and industry membership & \citet{hoberg2016text} \\
\texttt{revenue\_model}           & Business-model similarity              & \citet{hoberg2010product} \\
\texttt{value\_proposition}       & Product differentiation                & \citet{hoberg2010product} \\
\texttt{primary\_customers}       & Customer lead--lag links               & \citet{cohen2008economic} \\
\texttt{supply\_chain\_position}  & Production-network position            & \citet{Acemoglu2012,herskovic2018networks,barrot2016input} \\
\texttt{key\_competitors}         & Direct competitor links                & \citet{hoberg2016text} \\
\texttt{core\_technologies}       & Shared technology and R\&D exposure    & \citet{bybee2024business} \\
\texttt{strategic\_focus}         & Forward strategy and R\&D direction    & \citet{bybee2024business} \\
\texttt{geographic\_coverage}     & Common geographic and market exposure  & \citet{menzly2010market} \\
\texttt{financial\_profile}       & Size and growth characteristics        & \citet{harvey2016and} \\
\bottomrule
\end{tabular}
\end{table*}

\begin{table*}[h]
\centering
\caption{Each Distiller category maps to a mandatory disclosure surface in all five regulatory systems. References use each regulator's canonical section identifier; bracketed phrases give the local-language section title for the non-English systems.}
\label{tab:schema_grounding}
\scriptsize
\setlength{\tabcolsep}{3pt}
\begin{tabular}{@{}lccccc@{}}
\toprule
\textbf{Distiller category} & \textbf{SEC 10-K} & \textbf{JP \emph{yuho}} & \textbf{KR DART} & \textbf{TW MOPS} & \textbf{HKEx AR} \\
& (US) & (Japan) & (Korea) & (Taiwan) & (Hong Kong) \\
\midrule
\texttt{main\_business\_segments} & Item 1     & yuho Part~II.1 [\emph{Business}]      & II.1 [\emph{Business overview}]    & §3 [\emph{Business}]       & MD\&A: Business \\
\texttt{core\_technologies}       & Item 1     & yuho Part~II.2 [\emph{R\&D}]          & II.5 [\emph{R\&D}]                  & §5 [\emph{R\&D}]            & MD\&A: R\&D \\
\texttt{primary\_customers}       & Item 1     & yuho Part~II.1 [\emph{Customers}]     & II.1 [\emph{Major customers}]      & §3 [\emph{Customers}]      & MD\&A: Customers \\
\texttt{supply\_chain\_position}  & Item 1     & yuho Part~II.1 [\emph{Suppliers}]     & II.4 [\emph{Raw materials}]        & §4 [\emph{Raw materials}]  & MD\&A: Supply \\
\texttt{geographic\_coverage}     & Item 1     & yuho Part~I.3 [\emph{Establishments}] & I.4 [\emph{Sites}]                  & §3 [\emph{Operations}]      & MD\&A: Geog. \\
\texttt{financial\_profile}       & Item 6/8   & yuho Part~V [\emph{Financials}]       & III [\emph{Financials}]            & §6 [\emph{Financials}]     & Fin.\ Stmts \\
\texttt{revenue\_model}           & Item 7     & yuho Part~II.3 [\emph{Performance}]   & II.2 [\emph{Sales}]                & §3 [\emph{Revenue}]        & MD\&A: Rev. \\
\texttt{value\_proposition}       & Item 1     & yuho Part~II.1 [\emph{Strengths}]     & II.1 [\emph{Comp. advantage}]      & §3 [\emph{Comp. advantage}] & MD\&A: Strategy \\
\texttt{strategic\_focus}         & Item 7     & yuho Part~II.2 [\emph{Challenges}]    & II.6 [\emph{Strategy}]             & §5 [\emph{Strategy}]       & MD\&A: Outlook \\
\texttt{key\_competitors}         & Item 1     & yuho Part~II.1 [\emph{Competition}]   & II.1 [\emph{Competitors}]          & §3 [\emph{Competitors}]    & MD\&A: Comp. \\
\bottomrule
\end{tabular}
\end{table*}

\noindent \textbf{Practical use.}
\cref{tab:schema_grounding} records where each schema field is obtained across markets. Where a regulator uses sub-sections rather than a single dedicated section (e.g., \emph{primary customers} appears in both Part~II.1 and the risk-factor section of a Japanese \emph{yuho}), the Distiller prompt (\cref{app:distiller_prompt}) asks the model to merge the relevant fragments into a single concise English description. In the 200-filing audit, 89\% of human-audited fields are rated ``accurate'' (\cref{app:extraction_quality}), with the lowest source-language group at 82\%.

\end{document}